\documentclass[12pt]{article}

\usepackage{graphics}

\def\bfv{{\bf v}}
\def\bfu{{\bf u}}
\def\bfj{{\bf J}}
\def\bfnabla{{\bf \nabla}}

\begin{document}

\title{Evaporative Deposition Patterns Revisited: \\ Spatial Dimensions of the Deposit}
\author{Yuri O.\ Popov\thanks{Current address: Department of Physics, University of Michigan, 500 E.\ University Ave., Ann Arbor, MI 48109.  E-mail: {\tt yopopov@umich.edu}}}
\date{{\em Department of Physics, University of Chicago,} \\
{\em 5640 S.\ Ellis Ave., Chicago, IL 60637}}
\maketitle


\begin{abstract}
A model accounting for finite spatial dimensions of the deposit patterns in the evaporating sessile drops of colloidal solution on a plane substrate is proposed.  The model is based on the assumption that the solute particles occupy finite volume and hence these dimensions are of the steric origin.  Within this model, the geometrical characteristics of the deposition patterns are found as functions of the initial concentration of the solute, the initial geometry of the drop, and the time elapsed from the beginning of the drying process.  The model is solved analytically for small initial concentrations of the solute and numerically for arbitrary initial concentrations of the solute.  The agreement between our theoretical results and the experimental data is demonstrated, and it is shown that the observed dependence of the deposit dimensions on the experimental parameters can indeed be attributed to the finite dimensions of the solute particles.  These results are universal and do not depend on any free or fitting parameters; they are important for understanding the evaporative deposition and may be useful for creating controlled deposition patterns.
\end{abstract}

\begin{center}
\end{center}


\section{Introduction}

The problem of the so-called ``coffee-drop deposit'' has recently aroused great interest.  The residue left when coffee dries on the countertop is usually darkest and hence most concentrated along the perimeter of the stain.  Ring-like stains, with the solute segregated to the edge of a drying drop, are not particular to coffee.  Mineral rings left on washed glassware, banded deposits of salt on the sidewalk during winter, and enhanced edges in water color paintings are all examples of the variety of physical systems displaying similar behavior and understood by coffee-drop deposit terminology.

Understanding the process of drying of such solutions is important for many scientific and industrial applications, where ability to control the distribution of the solute during drying process is at stake.  For instance, in the paint industry, the pigment should be evenly dispersed after drying, and the segregation effects are highly undesirable.  Also, in the protein crystallography, attempts are made to assemble the two-dimensional crystals by using evaporation driven convection~\cite{pre1, pre2, dushkin}, and hence solute concentration gradients should be avoided.  On the other hand, in the production of nanowires~\cite{pre3} or in surface patterning~\cite{pre4} perimeter-concentrated deposits may be of advantage.  Recent important applications of this phenomenon related to DNA stretching in a flow have emerged as well~\cite{jpcb2}.  For instance, a high-throughput automatic DNA mapping was suggested~\cite{jpcb1}, where fluid flow induced by evaporation is used for both stretching DNA molecules and depositing them onto a substrate.  Droplet drying is also important in the attempts to create arrays of DNA spots for gene expression analysis.

Ring-like deposit patterns have been studied experimentally by a number of groups.  Difficulties of obtaining a uniform deposit~\cite{pre5}, deformation of sessile drops due to a sol-gel transition of the solute at the contact line~\cite{pre6, pre7}, stick-slip motion of the contact line of colloidal liquids~\cite{pre8, pre9}, multiple ring formation~\cite{shmuylovich}, and the effect of ring formation on the evaporation of the sessile drops~\cite{pre0} were all reported.  The evaporation of the sessile drops (regardless of the solute presence) has also been investigated extensively.  Constancy of the evaporation flux was demonstrated~\cite{jpcb3, jpcb4}, and the change of the geometrical characteristics (contact angle, drop height, contact-line radius) during drying was measured in detail~\cite{jpcb5, jpcb6, jpcb7, jpcb8}.

The most recent and complete experimental effort to date on coffee-drop deposits was conducted by Robert Deegan {\em et al.}~\cite{deegan1, deegan2, deegan3, deegan4}.  Most experimental data referred to in this work originate from observations and measurements of this group.  They reported extensive results on ring formation and demonstrated that these could be quantitatively accounted for.  The main ideas of the theory of solute transfer in such physical systems have also been developed in their work~\cite{deegan1}.  It was observed that the contact line of a drop of liquid remains pinned during most of the drying process.  While the highest evaporation occurs at the edges, the bulk of the solvent is concentrated closer to the center of the drop.  In order to replenish the liquid removed by evaporation at the edge, a flow from the inner to the outer regions must exist inside the drop.  This flow is capable of transferring all of the solute to the contact line and thus accounts for the strong contact-line concentration of the residue left after complete drying.  This theory is very robust since it is independent of the nature of the solute and only requires pinning of the edge during drying (which can occur in a number of possible ways: surface roughness, chemical heterogeneities {\em etc}).  Among other things, we will reproduce some of its results in this work.

Mathematically, the most complicated task is related to determining the evaporation rate from the surface of the drop.  An analogy between the diffusive concentration fields and the electrostatic potential fields was suggested~\cite{lebedev, jpcb0}, so that an equivalent electrostatic problem can be solved instead of the evaporation problem.  Important analytical solutions to this equivalent problem in various geometries were first derived by Lebedev~\cite{lebedev}, and a few useful consequences from these analytical results were later reported in Ref.~\cite{hu}.

In this work, we continue development of the theory of solute transfer and deposit growth.  Most previous works address the issue of the deposit mass accumulation at the drop boundary, however, they treat the solute particles as if they do not occupy any volume, and hence all the solute can be accommodated at the one-dimensional singularity of the contact line.  In reality, the solute deposit accumulated at the perimeter has some thickness, and the shape of the residue in a round drop resembles a ring rather than an infinitely thin circumference of the circle.  The earlier efforts were aimed at describing how the {\em mass\/} of the contact-line deposit grows with time and how it depends on such geometrical characteristics of the drop as its radius (for circular drops~\cite{deegan1, deegan2}) or its opening angle and the distance from the vertex (for pointed drops~\cite{popov1, popov2}).  Little attempt has been made to describe the geometrical characteristics of the contact-line deposit itself, for instance, the width and the height of the deposit ring.  At the same time, there is solid experimental data~\cite{deegan3, deegan4} on various geometrical characteristics of the ring and their dependence on time, the initial solute concentration, and the drop geometry.  Here we develop a simple model that addresses this lack of understanding of the geometrical properties of the contact-line deposit and accounts for the finite size of the deposit ring.  We attribute the finite volume of the deposit simply to the finite size of the solute particles, {\em i.e.}\ we assume the particles do occupy some volume and hence cannot be packed denser than certain concentration.  The model is solved in the simplest case of the circular geometry both analytically and numerically, and the results of the two methods are compared with the experimental data of Refs.~\cite{deegan3, deegan4} (and with each other).  It turns out that this model is sufficient to explain most of the collected data.  It should be noted that the model is as universal and robust (in its range of validity) as the zero-volume theory of Deegan {\em et al.}~\cite{deegan1, deegan2} since it is based on essentially the same physical principles.

The notion that the profile of the deposit could be found by the simple assumption that the solute becomes immobilized when the volume fraction reaches a threshold was originally suggested by Todd Dupont~\cite{dupont}.  First efforts to create a model were conducted by Robert Deegan~\cite{deegan3, deegan4} who formulated some physical assumptions, wrote them down mathematically, and obtained some early-time exponents.  Here we present the entire problem, including its full formulation and its analytical and numerical solutions (not reported previously).  In the next section, we formulate the model, describe the system, and address some issues of the geometry and the evaporation rate.  Then, we derive the governing equations from the conservation of mass and later solve them analytically for small initial concentrations of the solute and numerically for arbitrary initial concentrations of the solute.  A discussion section concludes this work.

\section{Model, assumptions, and geometry}

\subparagraph{System.}  We consider a sessile droplet of solution on a horizontal surface (substrate).  The nature of the solute is not essential for the mechanism.  The typical diameter of the solute particles in Deegan's experiments~\cite{deegan1, deegan2, deegan3, deegan4} was of the order of 0.1--1~$\mu$m; we will assume a similar order of magnitude throughout this work.  For smaller particles diffusion becomes important; for larger particles sedimentation may play an important role.

The droplet is bounded by the contact line in the plane of the substrate.  This (macroscopic) contact line is defined as the common one-dimensional boundary of all three phases (liquid, air and solid substrate).  We will restrict our attention to the case of the round drops, which is both of most practical importance and the easiest to treat mathematically.

We assume that the droplet is sufficiently small so that the surface tension is dominant, and the gravitational effects can be neglected.  Mathematically, the balance of the gravitational force and the surface tension is controlled by the ratio of the (maximal) hydrostatic pressure $\rho g h_{max}$ to the Laplace pressure $2 \sigma h_{max} / R_i^2$, where $\rho$ is the fluid density, $g$ is the gravitational constant, $\sigma$ is the surface tension at the liquid-air interface, $R_i$ is the drop radius in the plane of the substrate, and $h_{max}$ is the maximal height of the drop.  For the typical experimental conditions this ratio $\rho g R_i^2 / 2 \sigma$ is quite small (about 0.25), and thus gravity is indeed unimportant and the surface shape is governed mostly by the surface tension.  Our treatment will produce the main-order term in the expansion in this parameter, and since the parameter value is not an order of magnitude smaller than one, it may be necessary to construct the correctional terms for better quantitative agreement.  For the present purposes, even the main term turns out to be sufficient to obtain the agreement with the experimental results.

Experimentally, the contact line remains {\em pinned\/} during most of the drying process.  Therefore, we do {\em not\/} assume that the contact angle $\theta$ between the liquid-air interface and the plane of the substrate is constant in time.  A strongly pinned contact line can sustain a wide range of (macroscopic) contact angles.  The pinning mechanism can be described as self-pinning, {\em i.e.}\ pinning by the deposit brought to the contact line by the hydrodynamic flows caused by evaporation.  A pinned contact line entails fluid flow toward that contact line.  The ``elasticity'' of the liquid-air interface fixed at the contact line provides the force driving this flow.

We will deal with small contact angles ($\theta \ll 1$) as it is almost always the case in the experimental realizations, including the experiments of Ref.~\cite{deegan3} (typically, $\theta_{max} < 0.1$--0.3).  It will also be seen necessary to assume that the contact angle is small in order to obtain any analytical results in a closed form.  A drop with a small contact angle is necessarily {\em thin\/}, {\em i.e.}\ its maximal height is much smaller than its radius and the slope of the free surface is small ($|\bfnabla h| \ll 1$).  Thus, we consider small contact angles, or, equivalently, thin drops.

We also consider {\em slow\/} flows, {\em i.e.}\ flows with low Reynolds numbers, which amounts to the neglect of the inertial terms in the Navier-Stokes equation.

The free surface is described by the local mean curvature that is spatially uniform at any given moment of time, but changes with time as the droplet dries.  Ideally, the surface shape should be considered dynamically together with the flow field inside the drop.  However, as was shown earlier~\cite{popov2, popov4}, for flow velocities much lower than the characteristic velocity $v^* = \sigma/3\eta$ (where $\sigma$ is the surface tension and $\eta$ is the dynamic viscosity), which is about 24~m/s for water under normal conditions, one can consider the surface shape {\em independently\/} of the flow and use the equilibrium result at any given moment of time for finding the flow at that time.  Equivalently, the ratio of the viscous forces to the capillary forces is the capillary number $Ca = \eta \tilde v / \sigma$ (where $\tilde v$ is some characteristic value of the flow velocity, which is of the order of 1--10~$\mu$m/s), and this number is of the order of $10^{-8}$--$10^{-7}$ under typical experimental conditions.  Thus, the capillary forces are by far the dominant ones.

\subparagraph{Geometry and surface shape.}  The cylindrical coordinates $(r,\phi,z)$ will be used throughout this work, as they are most natural for the geometry of interest.  The origin is chosen in the center of the circular footprint of the drop on the substrate.  Coordinate $z$ is always normal to the substrate, and the substrate is described by $z = 0$, with $z$ being positive on the droplet side of the space.  Coordinates $(r,\phi)$ are the polar radius and the azimuthal angle, respectively, so that the contact line is described by $r = R_i$, where $R_i$ is the radius of the drop footprint.  Due to the axial symmetry of the problem and our choice of the coordinates, no quantity depends on the azimuthal angle $\phi$.

Our model pictures the drop as a two-component system (the components being ``the fluid'' and ``the solute''), which has two ``phases'': ``the liquid phase'' in the middle of the drop and ``the deposit phase'' near the contact line.  Both components are present in both phases, and the difference between the phases lies only in the concentration of the solute in each phase.  In the deposit phase, the volume fraction of the solute $p$ is high and {\em fixed\/} in both space and time.  Thus, $p$ is just a constant number, one can think of it as comparable to the close-packing fraction or unity.  (The case of $p = 1$ may seem to be special as there is no fluid in the deposit phase; however, for small initial concentrations of the solute this case will be seen to lead to exactly the same main order results.)  In the liquid phase, the volume fraction of the solute $\chi$ varies in space and changes with time, and it is relatively small compared to $p$.  The initial volume fraction $\chi_i = \chi(0)$ is constant throughout the drop; at later moments the solute gets redistributed due to the flows, and the concentration becomes different in different parts of the liquid phase.  The volume fraction of the fluid is then $(1-p)$ in the deposit phase and $(1-\chi)$ in the liquid phase.  Note that we do not require $\chi_i \ll p$ so far, although we do assume $\chi_i < p$.  It should also be emphasized that we do not presume there is any real ``phase difference'' between the so-called phases: one phase is just defined as having the maximal reachable solute fraction $p$ (the solute cannot move in this phase) while the other phase is characterized by lower solute fraction $\chi$ (in this phase the solute can move and hence its concentration can change in time and space).  Besides this difference, the phases are essentially identical.  The idea that the solute loses its mobility when its concentration exceeds some threshold was suggested by Todd Dupont~\cite{dupont}.

\begin{figure}
\begin{center}
\includegraphics{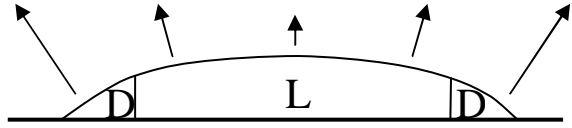}
\caption{Mutual location of the two ``phases'' in the drying drop:  L is the liquid phase, and D is the deposit phase.}
\label{phaseseps}
\end{center}
\end{figure}

\begin{figure}
\begin{center}
\includegraphics{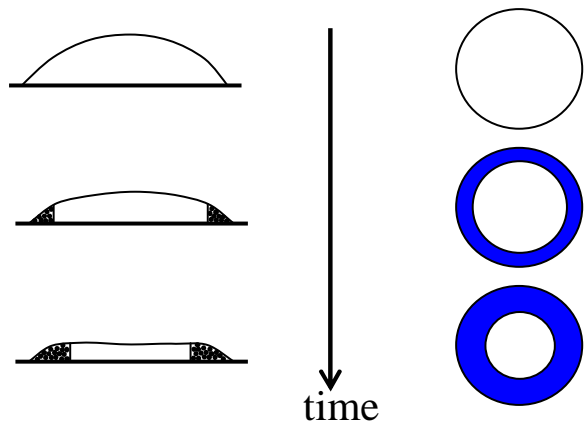}
\caption{Time evolution of the deposit phase growth: side view (left) and top view (right).  Only the deposit phase is shown.  Thickness of the ring is exaggerated compared to the typical experimental results.}
\label{evolutioneps}
\end{center}
\end{figure}

Since the drop is thin, we employ the vertically averaged flow velocity
\begin{equation}
\bfv(r) = \frac 1{h_t(r)} \int_0^{h_t(r)} \bfu_s(r,z) \, dz,
\label{defv}\end{equation}
where $\bfu_s(r,z)$ is the in-plane component of the local three-dimensional velocity $\bfu(r,z)$, and $h_t(r)$ is the thickness of the drop at distance $r$ from the center.  By making this approximation, we implicitly assume that there is no vertical segregation of the solute, and thus we turn our model into effectively two-dimensional.  This is done mostly for simplicity and is not expected to affect our main conclusions (see Discussion).  Within this model, it is natural to assume that the boundary between the phases is vertical.  Thus, the particles get stacked uniformly at all heights when being brought to the phase boundary by the hydrodynamic flow $\bfv(r)$.  This boundary can be pictured as a vertical wall at some radius $R(t)$ from the center of the drop, and this wall propagates from the contact line [located at $R_i = R(0)$] towards the center of the drop.  Fig.~\ref{phaseseps} illustrates the mutual location of the two phases, and Fig.~\ref{evolutioneps} schematically shows the time evolution of the drying process and growth of the deposit phase.

\begin{figure}
\begin{center}
\includegraphics{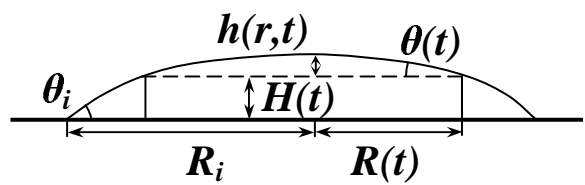}
\caption{Geometry of the problem.  Vertical scale is exaggerated in order to see the details, typically $H \ll R_i$ and $h \ll R_i$.}
\label{geometry-ring}
\end{center}
\end{figure}

The geometrical parameters of the model are shown in Fig.~\ref{geometry-ring}.  The radius of the drop is $R_i$, the radius of the phase boundary is $R(t)$, and $R(0) = R_i$.  The height of the phase boundary is $H(t)$, and the initial condition is $H(0) = 0$.  In the liquid phase, we conveniently split the total height of the free surface $h_t(r,t)$ into the sum of $H(t)$ and $h(r,t)$.

Since $H$ is independent of $r$, function $h(r)$ satisfies the Young-Laplace equation (the statement of the mechanical equilibrium of the liquid-air interface) 
\begin{equation}
2 K = - \frac{\Delta p}\sigma,
\label{laplace}\end{equation}
where $\sigma$ is the surface tension, $\Delta p$ is the pressure difference across the liquid-air interface, and $K$ is the mean curvature of the surface, uniquely related to the surface shape $h$ by differential geometry.  For typical drying conditions, $\Delta p$ and $h$ vary with time slowly.  As was shown earlier~\cite{popov2, popov4}, it is sufficient to find the equilibrium surface shape first, and then determine the velocity field for this fixed functional form of $h$ with time being just an adiabatic parameter, instead of solving for all the dynamical quantities simultaneously.  Thus, the right-hand side of Eq.~(\ref{laplace}) does {\em not\/} depend on the local coordinates of a point within the drop (although it does depend on time), and the equation itself expresses the global condition of spatial constancy of the mean curvature throughout the drop.  It defines the {\em equilibrium\/} surface shape at any given moment of time.

The solution to Eq.~(\ref{laplace}) with boundary condition $h(R) = 0$ is just a spherical cap, and hence the shape of the upper part of the drop (above the dashed line in Fig.~\ref{geometry-ring}) is just a spherical cap:
\begin{equation}
h(r,t) = \sqrt{\frac{R^2(t)}{\sin^2 \theta(t)} - r^2} - R(t) \cot\theta(t).
\label{sphericalcap}\end{equation}
Here $\theta(t)$ is the angle between the liquid-air interface and the substrate at phase boundary, and functions $R(t)$ and $\theta(t)$ are related via the right-hand side of Eq.~(\ref{laplace}):
\begin{equation}
R(t) = \frac{2\sigma}{\Delta p(t)} \sin\theta(t).
\end{equation}
In the limit of small contact angles, $\theta \ll 1$, the preceding expression adopts even simpler form:
\begin{equation}
h(r,t) = \frac{R^2(t) - r^2}{2R(t)} \theta(t) + O(\theta^3).
\label{h-finite}\end{equation}
Note that we do not assume that $\theta(t)$ and $h(r,t)$ are necessarily positive at all times: both can be negative at later drying stages, and the shape of the liquid-air interface may be concave.  Both convex and concave solutions for $h(r,t)$ are consistent with Eq.~(\ref{laplace}); the right-hand side of this equation can have either sign.  By definition, both $\theta(t)$ and $h(r,t)$ are positive when the surface is convex (and hence they are positive at the beginning of the drying process) and negative when the surface is concave.  The initial value of $\theta(t)$ coincides with the initial contact angle $\theta_i = \theta(0)$.

Clearly, there are three unknown functions of time in this geometry: $\theta(t)$, $R(t)$ and $H(t)$.  However, these quantities are not independent of each other.  Since we assume that the solute particles fill up the entire space between the substrate and the liquid-air interface when being brought to the phase boundary, the three geometrical functions are related by the constraint
\begin{equation}
\frac{dH}{dt} = - \theta \frac{dR}{dt}.
\label{constraint}\end{equation}
Physically, the angles between the liquid-air interface and the substrate are identical on both sides of the phase boundary ($\theta = |dH/dR|$), and hence $h(r)$ and its first derivative are continuous past this boundary.  Thus, there are actually only two {\em independent\/} functions of time, $\theta(t)$ and $R(t)$.  Condition~(\ref{constraint}) was first introduced by Robert Deegan~\cite{deegan3, deegan4}.

The geometrical definitions above allow one to determine the volume of each of the two phases.  Volume of the liquid phase is simply
\begin{equation}
V_L = \int_0^{R(t)} \left( h(r,t) + H(t) \right) \, 2\pi r dr = 2\pi \left( \frac{R^3 \theta}8 + \frac{R^2 H}2 \right) + O(\theta^3).
\end{equation}
Taking into account relation~(\ref{constraint}), an infinitesimal variation of this volume can be expressed via the infinitesimal variations of $\theta$ and $R$:
\begin{equation}
dV_L = \frac{\pi R^4}4 d\left( \frac{\theta}R \right) + 2\pi H R dR.
\label{dvl}\end{equation}
The first term is responsible for the motion of the liquid-air interface, and the second term corresponds to the inward shift of the phase boundary.  It is also straightforward to obtain an expression for the differential of the volume of the deposit phase, which has only the term related to the inward shift of the phase boundary:
\begin{equation}
dV_D = - 2\pi H R dR.
\label{dvd}\end{equation}
We will use the last two expressions in the following section.  We will also adopt the notation that subscripts L and D refer to the liquid and deposit phases, respectively.

Before proceeding to the main section, we will make a note on the evaporation rate.

\subparagraph{Evaporation rate.}  In order to determine the flow caused by evaporation, one needs to know the flux profile of liquid leaving each point of the surface.  This quantity is independent of the processes going on inside the drop and must be determined prior to considering any such processes.

The functional form of the evaporation rate $J(r)$ (defined as the evaporative mass loss per unit surface area per unit time) depends on the rate-limiting step, which can, in principle, be either the transfer rate across the liquid-vapor interface or the diffusive relaxation of the saturated vapor layer immediately above the drop.  We assume that the rate-limiting step is the diffusion of the saturated vapor.  Indeed, the transfer rate across the liquid-vapor interface is characterized by the time scale of the order of $10^{-10}$~s, while the diffusion process has characteristic times of the order of $R_i^2/D$ (where $D$ is the diffusion constant for vapor in air and $R_i$ is a characteristic size of the drop), which is of the order of seconds for water drops under typical drying conditions.  The diffusion-limited evaporation rapidly attains a steady state.  Indeed, the ratio of the time required for the vapor-phase water concentration to adjust to the changes in the droplet shape ($R^2/D$) to the droplet evaporation time $t_f$ is of the order of $(n_s - n_\infty)/\rho \approx 10^{-5}$, where $n_s$ is the density of the saturated vapor just above the liquid-air interface, $n_\infty$ is the ambient vapor density, and $\rho$ is the fluid density~\cite{hu}, {\em i.e.}\ the vapor concentration adjusts rapidly compared to the evaporation time.

As the rate-limiting process is the diffusion, vapor density $n$ above the liquid-vapor interface obeys the diffusion equation.  Since the process is quasi-steady, this diffusion equation reduces to the Laplace equation
\begin{equation}
\nabla^2 n = 0.
\end{equation}
This equation is to be solved together with the following boundary conditions: (a) along the surface of the drop the air is saturated with vapor and hence $n$ at the interface is the constant density of the saturated vapor $n_s$, (b) far away from the drop the density approaches the constant ambient vapor density $n_\infty$, and (c) the vapor cannot penetrate the substrate and hence $\partial_z n = 0$ at the substrate outside of the drop.  Having found the vapor density, one can obtain the evaporation rate $\bfj = - D \bfnabla n$, where $D$ is the diffusion constant.

\begin{figure}
\begin{center}
\includegraphics{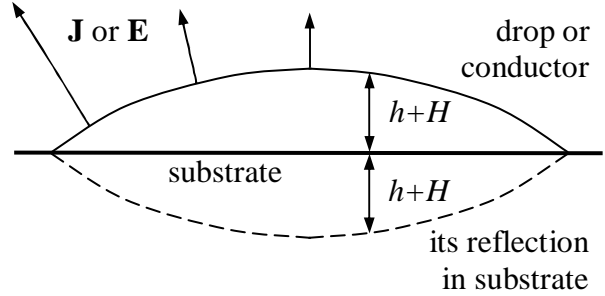}
\caption{Illustration of the analogy between the evaporation rate {\bf J} for a liquid drop and the electric field {\bf E} for a conductor.  Consideration of the drop (or conductor) and its reflection in the plane of the substrate significantly simplifies the boundary problem.}
\label{evaprateeps}
\end{center}
\end{figure}

This boundary problem is mathematically equivalent to that of a charged conductor of the same geometry at constant potential if we identify $n$ with the electrostatic potential and $\bfj$ with the electric field.  Moreover, since there is no component of $\bfj$ normal to the substrate, we can further simplify the boundary problem by considering a conductor of the shape of our drop plus its reflection in the plane of the substrate in the full space instead of viewing only the semi-infinite space bounded by the substrate (Fig.~\ref{evaprateeps}).  This reduces the number of boundary conditions to only two: (a) $n = n_s$ on the surface of the conductor, and (b) $n = n_\infty$ at infinity.  The shape of the conductor (the drop and its reflection in the substrate) is now symmetric with respect to the plane of the substrate and resembles a symmetrical double-convex lens comprised of two spherical caps.  This equivalent electrostatic problem of finding the electric field around the conductor at constant potential in the infinite space is much simpler than the original problem in the semi-infinite space.  The reflection technique for finding the evaporation field on the basis of the analogy between the diffusion and the electrostatics was originally used by Deegan {\em et al.}~\cite{deegan1, deegan2}.

Even in the circular geometry the equivalent problem is still quite complicated despite the visible simplicity.  We consider an object whose symmetry does not match the symmetry of any simple orthogonal coordinate system of the three-dimensional space.  In order to solve the Laplace equation, one is forced to introduce a special coordinate system (the so-called toroidal coordinates) with heavy use of the special functions.  The full solution to this problem is provided in the Appendix.

\begin{figure}
\begin{center}
\includegraphics{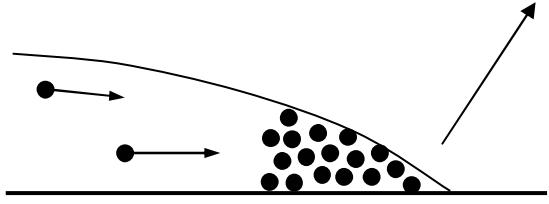}
\caption{Presence of particles in the deposit does not obstruct fluid evaporation at the edge of the drop.  All the necessary fluid is supplied, and it is this motion of the fluid that brings the particles to the deposit phase.  Also shown schematically is the fact that the boundary between the phases is vertical and the particles get stacked at full height between the substrate and the free surface of the drop.}
\label{suckseps}
\end{center}
\end{figure}

The evaporation rate depends only on the overall shape of the drop, and evaporation occurs in the same fashion from both phases.  We assume that the evaporation is not influenced by any motion of the solute inside the drop, and the necessary amount of fluid can always be supplied to the regions of the highest evaporation near the contact line.  Physically, high evaporation near the edge is what brings the solute to the contact line, and we assume that presence of the deposit does not obstruct the motion of the fluid (Fig.~\ref{suckseps}).  Since the drop is thin and the contact angle is small, we will use expression
\begin{equation}
J(r) = \frac{2}\pi \frac{D (n_s - n_\infty)}{\sqrt{R_i^2 - r^2}}
\label{evaprate-circular}\end{equation}
for the evaporation rate (derived in the Appendix for no-solute drops in the limit $\theta \ll 1$), which has the one-over-the-square-root divergence near the contact line intuitively expected from the electrostatics.  The real situation may be different from the assumed above when $p$ is large or comparable to 1, and the edge of the area where the evaporation occurs may be located near the boundary of the phases instead of the contact line.  However, for small initial concentrations of the solute, the main order result will be insensitive to the exact location of the singularity of the evaporation rate: whether it is located at the contact line or near the boundary of the phases.  We will further comment on this case of the ``dry deposit'' when we obtain the full system of equations.

\section{Principal equations}

\subparagraph{Global conservation of fluid.}  The essence of the entire theory can best be summarized in one sentence: ``It is all about the conservation of mass.''  Indeed, as we will see by the end of this section, all three governing equations obtained here represent the conservation of mass (or volume) in one form or another.

We start from the global conservation of fluid in the drop.  Since the amount of solute within the drop does not change during the drying process, the change of the entire drop volume is equal to the change of the amount of fluid.  This fluid gets evaporated from the surface, and the total change of the fluid volume equals to the amount evaporated from the surface:
\begin{equation}
\left. dV \right|_{tot} = \left. dV^F \right|_{surf}.
\label{aux1}\end{equation}
By convention, superscripts F and S refer to the fluid and the solute components, respectively (while subscripts L and D continue to denote phases).  The total change of the drop volume is the sum of the volume changes of each phase:
\begin{equation}
\left. dV \right|_{tot} = dV_L + dV_D = \frac{\pi R^4}4 d\left( \frac{\theta}R \right),
\label{aux2}\end{equation}
where $dV_L$ and $dV_D$ were found in the preceding section [Eqs.~(\ref{dvl}) and (\ref{dvd})].  The volume of fluid evaporated from the surface can be determined from the known evaporation rate:
\begin{equation}
\left. \frac{dV^F}{dt} \right|_{surf} = - \int_0^{R_i} \frac{J(r)}\rho \sqrt{1 + (\partial_r h)^2} \, 2\pi r dr = - \frac{4 D (n_s - n_\infty) R_i}\rho,
\label{aux3}\end{equation}
where $\rho$ is the fluid density.  We neglected the gradient of $h(r)$ with respect to unity (which is always legitimate for thin drops) and used $J(r)$ of Eq.~(\ref{evaprate-circular}).  Thus, Eqs.~(\ref{aux1}), (\ref{aux2}), and (\ref{aux3}) yield the first main differential equation of this section:
\begin{equation}
R^4 \frac{d}{dt}\left(\frac\theta{R}\right) = - \frac{16 D (n_s - n_\infty) R_i}{\pi\rho}.
\label{main1}\end{equation}
This equation represents the global conservation of fluid in the drop and relates the time dependencies of $\theta(t)$ and $R(t)$.

\begin{figure}
\begin{center}
\includegraphics{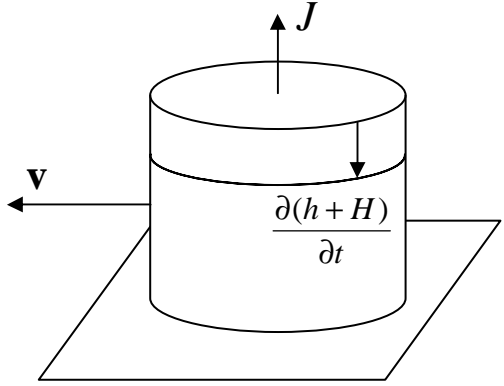}
\caption{Conservation of mass: the liquid-vapor interface lowers exactly by the amount of fluid evaporated from the surface plus the difference between the outflow and the influx of fluid from the adjacent regions.}
\label{consmasseps}
\end{center}
\end{figure}

\subparagraph{Local conservation of mass.}  The next equation represents the {\em local\/} conservation of mass.  There are two components in the liquid phase, and hence we write a separate equation for each of them.  Since a free particle of the appropriate size reaches the speed of the flow in about 50~ns in water under normal conditions~\cite{popov3}, the solute particles are simply carried along by the flow, and the velocities of each component are identical at each point within the liquid phase [and equal to the depth-averaged fluid velocity $\bfv$ defined in Eq.~(\ref{defv})].  The local conservation of {\em fluid\/} can be written in the form:
\begin{equation}
\bfnabla\cdot\left[(1 - \chi)(h + H)\bfv\right] + \frac{J}{\rho}\sqrt{1+(\nabla h)^2} + \partial_t \left[(1 - \chi)(h + H)\right] = 0,
\label{cons-fluid}\end{equation}
where $\chi$ is the volume fraction of solute at a given point within the liquid phase, and each of the quantities $(h + H)$, $J$, $\chi$, and $\bfv$ is a function of distance $r$ and time $t$.  [We drop the $(\nabla h)^2$ part of the second term everywhere in this work since it is always small compared to unity for small contact angles.]  This equation represents the fact that the rate of change of the fluid amount in a volume element (column) above an infinitesimal area on the substrate (third term) is equal to the negative of the sum of the net flux of fluid out of the column (first term) and the amount of fluid evaporated from the surface element on top of that column (second term); Fig.~\ref{consmasseps} illustrates the idea.  A similar equation can also be written for the local conservation of {\em solute\/}, but without the evaporation term:
\begin{equation}
\bfnabla\cdot\left[\chi(h + H)\bfv\right] + \partial_t \left[\chi(h + H)\right] = 0.
\label{cons-solute}\end{equation}
Adding the two equations and employing the linearity of the differential operations, one obtains:
\begin{equation}
\bfnabla\cdot[(h + H)\bfv] + \frac{J}{\rho} + \partial_t (h + H) = 0.
\label{cons-vol}\end{equation}
This relation could have been obtained if we considered only one component with volume fraction 1 in the liquid phase, and this equivalence should be of no surprise: when the solute moves in exactly the same fashion as the fluid does, any differentiation between the two is completely lost (from the point of view of the conservation of volume).  Note that if evaporation were too intensive, this equivalence would not hold, as there might be an insufficient amount of fluid coming into a volume element, and the solution could get completely dry (only the solute component would be left).  We implicitly assume this is not the case for our liquid phase where the solute fraction is relatively small and the evaporation is not too strong.

In circular geometry, due to the symmetry, the flow is radial and independent of $\phi$.  Thus, Eq.~(\ref{cons-vol}) can be resolved with respect to the radial component of the velocity:
\begin{equation}
v_r(r,t) = - \frac{1}{r (h + H)} \int _0^r \left( \frac{J}\rho + \partial_t h + \partial_t H \right) \, r dr.
\end{equation}
Straightforward integration with $h(r,t)$ of Eq.~(\ref{h-finite}), $dH/dt$ of Eq.~(\ref{constraint}), and $J(r)$ of Eq.~(\ref{evaprate-circular}) and employment of Eq.~(\ref{main1}) for $d(\theta/R)/dt$ yield
\begin{equation}
v_r(r,t) = \frac{2 D (n_s - n_\infty)}{\pi\rho} \frac{R_i}r \frac{\sqrt{1 - \left( \frac{r}{R_i} \right)^2} - \left[ 1 - \left( \frac{r}R \right)^2 \right]^2}{\frac{R \theta}2 \left[ 1 - \left( \frac{r}R \right)^2 \right] + H}.
\label{vr-finite}\end{equation}
This expression for the flow velocity at each point $r$ within the liquid phase in terms of the time-dependent geometrical characteristics of the drop $\theta(t)$, $R(t)$, and $H(t)$ is a direct consequence of the local conservation of mass.

With the velocity in hand, we can compute the time it takes an element of fluid initially located at distance $r_i$ from the center to reach the contact line.  First, only the particles initially located near the contact line reach that contact line.  As time goes by, the particles initially located further away from the contact line and in the inner parts of the drop reach the contact line.  Finally, the particles initially located in the innermost parts of the drop ({\em i.e.}\ near its center) reach the contact line as well.  The more time elapsed, the more particles reached the contact line and the larger the area is where they were spread around initially.  One can view this process as inward propagation of the inner boundary of the set of the initial locations of the particles that have reached the contact line by time $t$.  As is easy to understand, the velocity of this front is equal to the negative of the vector of the fluid velocity at each point (the fluid and the particles move together towards the contact line while this front moves away from it, hence a minus sign).  We label $r_i(t)$ the initial location of the solute particles that reach the phase boundary (and become part of the deposit ring) at time $t$.  Since the solute particles from the outer areas of the drop reach the deposit phase sooner than the particles from the inner areas, this function is monotonically decreasing, and its derivative is simply related to $v_r$ found in the preceding paragraph [Eq.~(\ref{vr-finite})]:
\begin{equation}
\frac{dr_i}{dt} = - v_r(r_i,t).
\end{equation}
Thus, the second principal equation of this section is
\begin{equation}
\frac{dr_i}{dt} = - \frac{2 D (n_s - n_\infty)}{\pi\rho} \frac{R_i}{r_i} \frac{\sqrt{1 - \left( \frac{r_i}{R_i} \right)^2} - \left[ 1 - \left( \frac{r_i}R \right)^2 \right]^2}{\frac{R \theta}2 \left[ 1 - \left( \frac{r_i}R \right)^2 \right] + H}.
\label{main2}\end{equation}
This equation relates $r_i(t)$ to the time dependencies of the geometrical parameters of the drop [$\theta(t)$, $R(t)$, and $H(t)$].

\subparagraph{Global conservation of solute.}  The volume of solute in the deposit phase $V_D^S$ at time $t$ is equal to the volume of solute outside the circle of radius $r_i(t)$ at time 0 (since all the solute between $r_i(t)$ and $R_i$ becomes part of the deposit by time $t$).  The latter volume can be found by integrating $h(r,0)$ over the area swept by the fluid on its way from $r_i$ to the contact line and multiplying the result by the initial volume fraction of solute $\chi_i$:
\begin{equation}
V_D^S = \chi_i \int_{r_i}^{R_i} h(r,0) \, 2\pi r dr = V^S \left[ 1 - \left( \frac{r_i}{R_i} \right)^2 \right]^2,
\label{vds}\end{equation}
where $V^S = \pi \chi_i R_i^3 \theta_i / 4$ is the total volume of solute in the drop.  On the other hand, the volume of solute in the deposit phase is just the constant fraction $p$ of the volume of the entire deposit phase:
\begin{equation}
V_D^S = p V_D.
\end{equation}
Equating the right-hand sides of these two equations, taking the time derivatives of both sides, and making use of the already determined $dV_D$ of Eq.~(\ref{dvd}), we obtain the third principal equation of this section:
\begin{equation}
\chi_i R_i^3 \theta_i \left[ 1 - \left( \frac{r_i}{R_i} \right)^2 \right] \frac{d}{dt} \left[ 1 - \left( \frac{r_i}{R_i} \right)^2 \right] = - 4 p H R \frac{dR}{dt}.
\label{main3}\end{equation}
This equation represents the global conservation of solute in the drop.

Thus, we have four unknown functions of time: $\theta(t)$, $R(t)$, $H(t)$, and $r_i(t)$, and four independent differential equations for these functions: Eqs.~(\ref{constraint}), (\ref{main1}), (\ref{main2}), and (\ref{main3}).  In reality, we need only three of these functions: $\theta(t)$, $R(t)$, and $H(t)$; however, there is no simple way to eliminate $r_i(t)$ from the full system and reduce the number of equations.  Having solved this system of equations, we will be able to fully characterize the dimensions of the deposit phase and describe the evolution of the deposit ring.  The following section is devoted to the details and the results of this solution.

Here we will only comment on how this system changes in the case of the completely dry solute $p = 1$.  In this case there is no evaporation from the surface of the deposit phase, and the effective edge of the evaporating area is somewhere in the vicinity of the phase boundary.  Assuming the same one-over-the-square-root divergence of the evaporation rate at $r = R$ instead of $r = R_i$ [which mathematically means substitution of $R$ in place of $R_i$ in Eq.~(\ref{evaprate-circular})] and conducting a derivation along the lines of this section, one can obtain a very similar system of four differential equations.  These equations would be different from Eqs.~(\ref{constraint}), (\ref{main1}), (\ref{main2}), and (\ref{main3}) in only two minor details.  First, Eqs.~(\ref{main1}) and (\ref{main2}) would lose all indices $i$ at all occasions of $R_i$ ({\em i.e.}\ one should substitute $R$ for all $R_i$ in both equations).  Second, $p$ should be set to 1 in Eq.~(\ref{main3}).  Apart from these details, the two systems would be identical.  As we will see in the following section, this difference between the two systems is not important in the main order in a small parameter introduced below, and thus this ``dry-solute'' case does not require any special treatment contrary to the intuitive prudence.

\section{Results}

\subparagraph{Analytical results in the limit of small initial concentrations of the solute.}  So far we have not introduced any small parameters other than the initial contact angle $\theta_i \ll 1$.  In particular, equations~(\ref{constraint}), (\ref{main1}), (\ref{main2}), and (\ref{main3}) were obtained without assuming any relation between $p$ and $\chi_i$ other than the non-restrictive condition $\chi_i < p$.  In order to find the analytical solution to this system, we will have to assume that $\chi_i \ll p$.  Then, we will solve the same system of differential equations numerically for an arbitrary relation between $\chi_i$ and $p$.

Assumption $\chi_i \ll p$ physically means that the solute concentration in the liquid phase is small --- it is much smaller than the concentration of close packing or any other comparable number of the order of 1.  This is the case for most practical realizations of the ring deposits in experiments and observations: the solute concentration rarely exceeds 10\% of volume, and in most cases it is far lower.  If the volume fraction of the solute is small, then the solute volume is also small compared to the volume of the entire drop.  Hence, the deposit phase, which consists mostly of the solute, must also have small volume compared to the volume of the entire drop.  Thus, if the initial volume fraction $\chi_i$ is small, then the dimensions of the deposit ring must be small compared to the corresponding dimensions of the entire drop.

Let us now introduce parameter $\epsilon$ that is small when $\chi_i/p$ is small.  We do not fix its functional dependence on $\chi_i/p$ for the moment:
\begin{equation}
\epsilon = f\left(\frac{\chi_i}p\right) \ll 1,
\label{epsilon-def}\end{equation}
where $f$ is an arbitrary increasing function of its argument.  Then we postulate that the ring width is proportional to this parameter:
\begin{equation}
R(t) = R_i \left[ 1 - \epsilon \tilde W(t) \right],
\label{tilde-w-def}\end{equation}
where $\tilde W(t)$ is an arbitrary dimensionless function and we explicitly introduced the dimensionality via $R_i$.  Obviously, $\tilde W(0) = 0$.  So far we simply wrote mathematically that the ring width is small whenever the initial volume fraction of the solute is small.  Next, we introduce a dimensionless variable for the angle $\theta(t)$:
\begin{equation}
\theta(t) = \theta_i \tilde\theta(t),
\label{tilde-theta-def}\end{equation}
where both $\theta(t)$ and $\theta_i$ are small, while the newly introduced function $\tilde\theta(t)$ is arbitrary [in particular, $\tilde\theta(0) = 1$].  Due to the geometrical constraint~(\ref{constraint}), the height of the ring $H(t)$ must be linear in small parameters $\epsilon$ and $\theta_i$ and directly proportional to the only dimensional scale $R_i$:
\begin{equation}
H(t) = \epsilon \theta_i R_i \tilde H(t),
\label{tilde-h-def}\end{equation}
where $\tilde H(t)$ is yet another dimensionless function of time [$\tilde H(0) = 0$], related to functions $\tilde W(t)$ and $\tilde\theta(t)$ by an expression similar to Eq.~(\ref{constraint}).  The last dimensionless variable is introduced in place of the fourth unknown function $r_i(t)$:
\begin{equation}
\tilde V(t) = 1 - \left( \frac{r_i(t)}{R_i} \right)^2,
\label{tilde-v-def}\end{equation}
with the initial condition $\tilde V(0) = 0$.  Thus, we introduced four new dimensionless variables in place of the four original ones and explicitly separated their dependence on small parameters $\epsilon$ and $\theta_i$.  Finally, we define the dimensionless time $\tau$ as:
\begin{equation}
\tau = \frac{t}{t_f},
\label{tau}\end{equation}
where $t_f$ is a combination of system parameters with the dimensionality of time:
\begin{equation}
t_f = \frac{\pi \rho R_i^2 \theta_i}{16 D (n_s - n_\infty)}.
\label{t-f}\end{equation}
In the limit $\chi_i/p \to 0$ this combination represents the time at which all the solute reaches the deposit phase; for finite $\chi_i/p$ it does not have so simple interpretation.

Substitution of all the definitions of the preceding paragraph into the original system of equations~(\ref{constraint}), (\ref{main1}), (\ref{main2}), and (\ref{main3}) and retention of only the leading and the first correctional terms in $\epsilon$ yield the following simplified system of equations:
\begin{equation}
\frac{d\tilde H}{d\tau} = \tilde\theta \frac{d\tilde W}{d\tau},
\label{dimless1}\end{equation}
\begin{equation}
\frac{d\tilde\theta}{d\tau} + \epsilon \tilde\theta \frac{d\tilde W}{d\tau} - 3 \epsilon \tilde W \frac{d\tilde\theta}{d\tau} = -1,
\label{dimless2}\end{equation}
\begin{equation}
\frac{d\tilde V}{d\tau} = \frac{\sqrt{\tilde V} - \tilde V^2 \left[ 1 - 4 \epsilon \tilde W \left(\tilde V^{-1} - 1\right) \right]}{2 \tilde\theta \tilde V \left[ 1 - \epsilon \tilde W \left(2\tilde V^{-1} - 1\right) \right] + 4 \epsilon \tilde H},
\label{dimless3}\end{equation}
\begin{equation}
\frac{\chi_i}p \tilde V \frac{d\tilde V}{d\tau} = 4 \epsilon^2 \tilde H \frac{d\tilde W}{d\tau} \left(1 - \epsilon \tilde W\right).
\label{dimless4}\end{equation}
As is apparent from the last equation, parameter $\epsilon^2$ {\em must\/} be proportional to $\chi_i/p$.  Since the separation of the ring width into $\epsilon$ and $\tilde W$ in Eq.~(\ref{tilde-w-def}) is absolutely arbitrary, parameter $\epsilon$ is defined up to a constant multiplicative factor.  Therefore, we {\em set\/} this factor in such a way that $\epsilon^2$ is {\em equal\/} to $\chi_i/p$:
\begin{equation}
\epsilon = \sqrt{\frac{\chi_i}p}.
\end{equation}
This fixes the function $f$ from the original definition~(\ref{epsilon-def}).

The differential equations in the system~(\ref{dimless1})--(\ref{dimless4}) are still coupled.  However, in the main (zeroth) order in $\epsilon$, the equations clearly {\em decouple\/}: the second equation can be solved with respect to $\tilde\theta(\tau)$ independently of all the others, then the third equation can be solved with respect to $\tilde V(\tau)$ independently of the first and the fourth, and finally the first and the fourth equations can be solved together as well.  Thus, one can obtain the following main-order solution to the system of equations above with the appropriate initial conditions:
\begin{equation}
\tilde\theta(\tau) = 1 - \tau,
\label{tilde-theta-res}\end{equation}
\begin{equation}
\tilde V(\tau) = \left[1 - (1 - \tau)^{3/4}\right]^{2/3},
\label{tilde-v-res}\end{equation}
\begin{equation}
\tilde H(\tau) = \sqrt{\frac{1}3 \left[ {\rm B}\left(\frac{7}3,\frac{4}3\right) - {\rm B}_{(1 - \tau)^{3/4}}\left(\frac{7}3,\frac{4}3\right) \right]},
\label{tilde-h-res}\end{equation}
\begin{equation}
\tilde W(\tau) = \int_0^\tau \frac{1}{8 \tilde H(\tau')} \frac{\left[ 1 - (1 - \tau')^{3/4} \right]^{1/3}}{(1 - \tau')^{1/4}} \, d\tau'.
\label{tilde-w-res}\end{equation}
Here ${\rm B}(a,b) = \int_0^1 x^{a-1} (1-x)^{b-1} \, dx$ is the complete beta-function, ${\rm B}_z(a,b) = \int_0^z x^{a-1} (1-x)^{b-1} \, dx$ is the incomplete beta-function ($a > 0$, $b > 0$, and $0 \le z \le 1$), and the integral in the last equation cannot be expressed in terms of the standard elementary or special functions.  In a similar fashion, systems of equations of the higher orders in $\epsilon$ can be written [only the first-order corrections are kept in the system~(\ref{dimless1})--(\ref{dimless4})], and the higher-order terms can also be constructed  up to an arbitrary order.

A system of equations similar to our system~(\ref{dimless1})--(\ref{dimless4}) was presented by Robert Deegan in works~\cite{deegan3, deegan4}.  However, some terms of the first order in concentration were missing and no analytical solution to the system of equations was obtained in those works.  Here we derive the equations in a systematic way.  The analytical solution~(\ref{tilde-theta-res})--(\ref{tilde-w-res}) is provided for the first time.

\begin{figure}
\begin{center}
\includegraphics{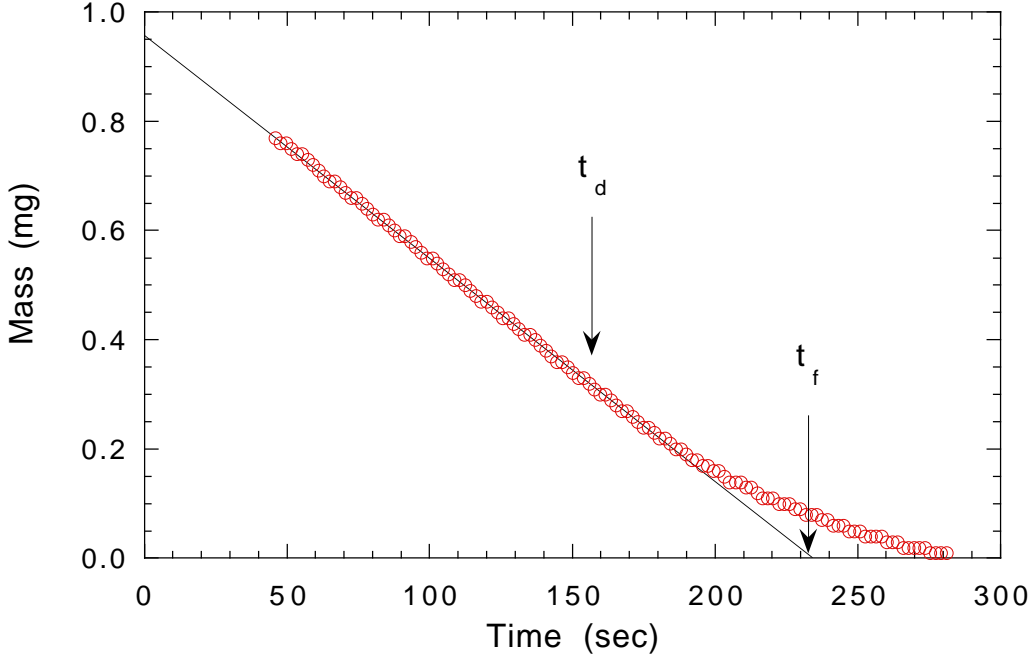}
\caption{Mass of a drying drop as a function of time.  Experimental results, after Refs.~\cite{deegan3, deegan4}.  The line running through the data is a linear fit.  (Courtesy Robert Deegan.)}
\label{thetatimeeps}
\end{center}
\end{figure}

\begin{figure}
\begin{center}
\includegraphics{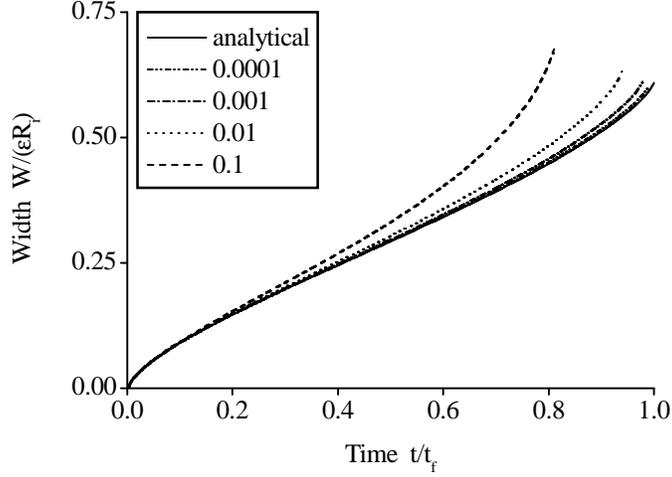}

(d)

\caption{Results: dependence of the geometrical characteristics of the drop on time $t$.  In each plot, the solid curve is the analytical result in the limit $\chi_i/p \to 0$, while the other curves are the numerical results.  Different numerical curves correspond to different initial concentrations of the solute; values of parameter $\chi_i/p$ are shown at each curve.  (a) Angle $\theta$ between the liquid-air interface and the substrate at the phase boundary.  (b) Volume fraction of the solute in the deposit phase $V_D^S/V^S$.  (c) Height of the phase boundary $H$ (in units of $\theta_i R_i \sqrt{\chi_i/p}$).  (d) Width of the deposit ring $W$ (in units of $R_i \sqrt{\chi_i/p}$).}
\label{resultseps}
\end{center}
\end{figure}

How do our results~(\ref{tilde-theta-res})--(\ref{tilde-w-res}) translate into the original variables?  The first two of them [Eqs.~(\ref{tilde-theta-res}) and (\ref{tilde-v-res})] reproduce earlier results.  In terms of the dimensional variables Eq.~(\ref{tilde-theta-res}) represents the linear decrease of the angle between the liquid-air interface and the substrate at the phase boundary with time:
\begin{equation}
\theta(t) = \theta_i \left(1 - \frac{t}{t_f}\right)
\label{theta-result}\end{equation}
[plotted by the solid line in Fig.~\ref{resultseps}(a)].  This is a direct analog of Eq.~(\ref{theta-circular}) for the contact angle in the no-solute case, as is clear from the definition of $t_f$ [Eq.~(\ref{t-f})].  So, angle $\theta$ in the case of the finite-volume solute depends on time in exactly the same fashion as the contact angle in the no-solute case does.  This expression also provides an interpretation of $t_f$: it is the time at which the free surface of the liquid phase becomes flat.  Before $t_f$ this surface is convex, after $t_f$ it becomes concave and bows inward (until it touches the substrate).  Thus, $t_f$ is generally {\em not\/} the total drying time.  In the limit $\chi_i/p \to 0$ the height of the deposit ring is going to zero and the two times are the same.  For finite values of this parameter the total drying time is longer than the time at which the liquid-air interface becomes flat.  Eq.~(\ref{theta-result}) has been verified in the experiments~\cite{deegan3, deegan4} where the mass of the drop was measured as a function of time (Fig.~\ref{thetatimeeps}).  Since the mass of a thin drop is directly proportional to $\theta$, these results confirm the linearity of $\theta(t)$ during most of the drying process.

The second equation~(\ref{tilde-v-res}) has a direct analog in the case of the zero-size solute particles.  In the original variables it can be rewritten as
\begin{equation}
\left( 1 - \frac{t}{t_f} \right)^{3/4} + \left[ 1 - \left(\frac{r_i(t)}{R_i}\right)^2 \right]^{3/2} = 1,
\label{ri-result}\end{equation}
which is identical to Eq.~(3.24) of Ref.~\cite{popov4} obtained for the zero-volume solute.  Clearly, $r_i = R_i$ when $t = 0$, and $r_i = 0$ when $t = t_f$.  According to Eqs.~(\ref{vds}) and (\ref{tilde-v-def}), the fraction of solute in the deposit phase $V_D^S / V^S$ is 
\begin{equation}
\frac{V_D^S}{V^S} = \tilde V^2 = \left[1 - \left( 1 - \frac{t}{t_f} \right)^{3/4} \right]^{4/3}
\label{fraction-result}\end{equation}
[plotted by the solid line in Fig.~\ref{resultseps}(b)].  This fraction is 0 at $t = 0$ and becomes 1 at $t = t_f$.  Thus, $t_f$ can also be interpreted as the time at which all the solute particles become part of the deposit phase.  So far, the results of this finite-volume model coincide with the results of the zero-volume case considered earlier~\cite{deegan1, deegan2}.

\begin{figure}
\begin{center}
\includegraphics{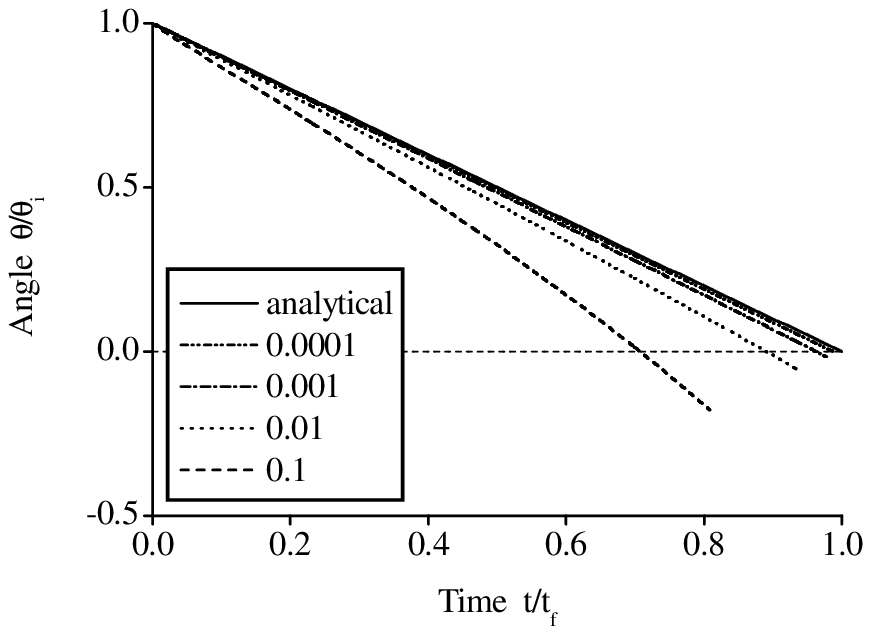}

(a)

\includegraphics{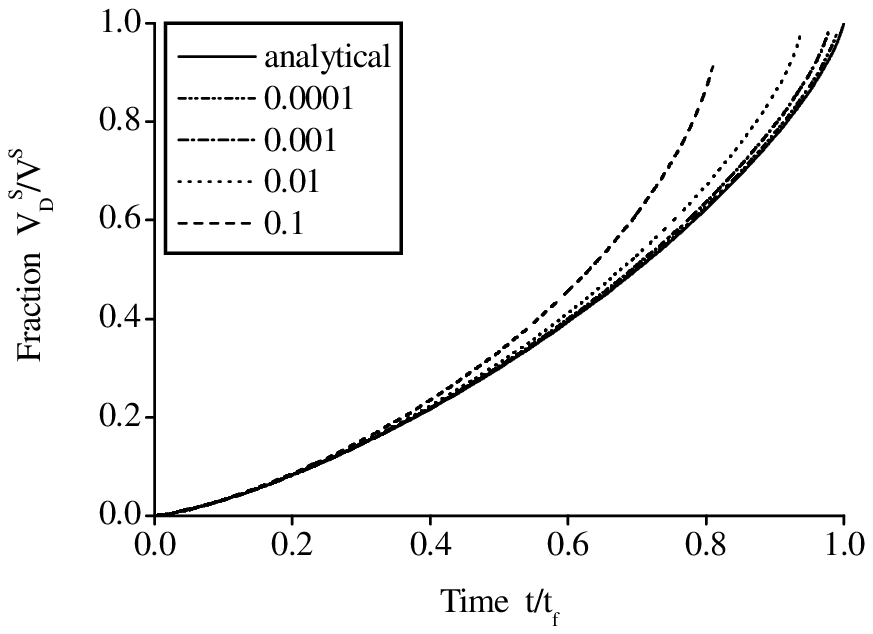}

(b)

\includegraphics{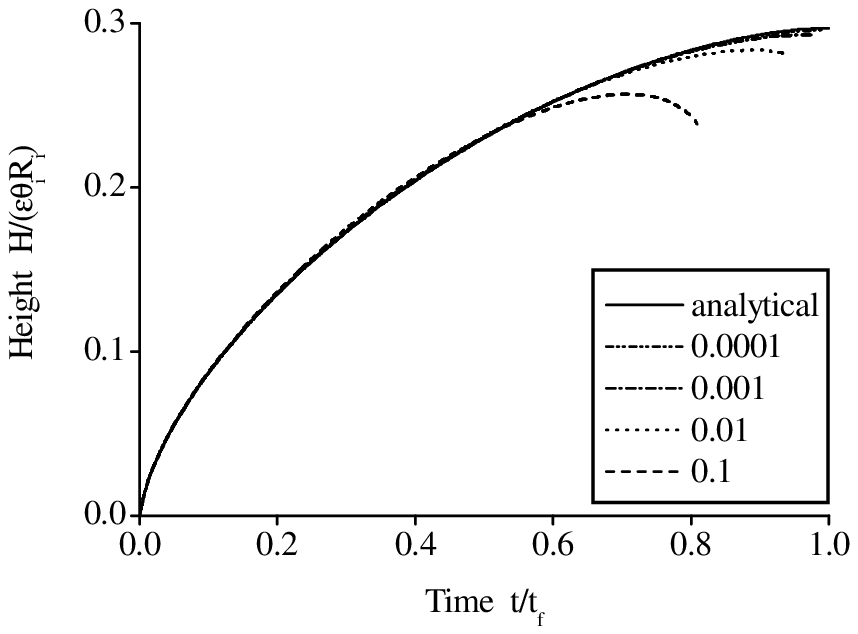}

(c)

\end{center}
\end{figure}


\begin{figure}
\begin{center}
\includegraphics{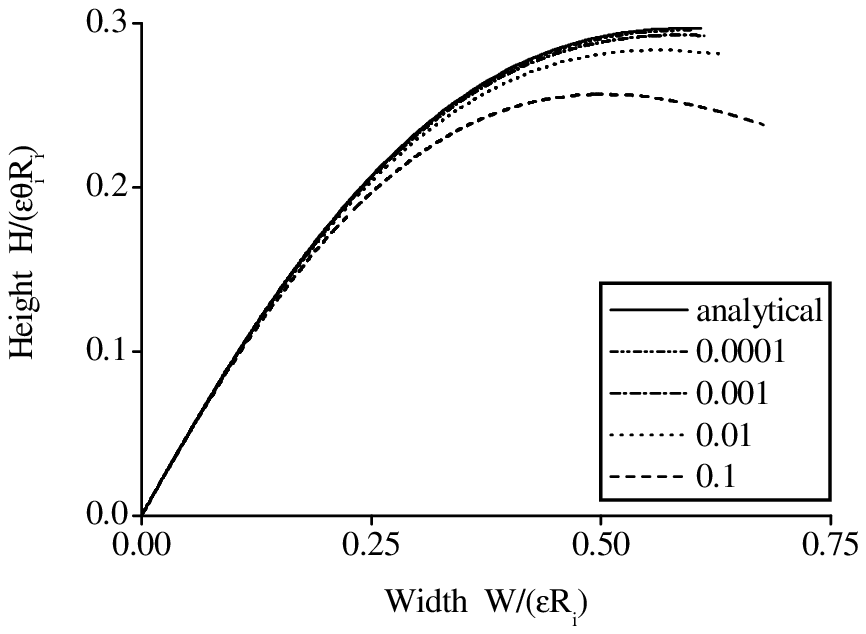}
\caption{Deposit ring profile: dependence of the height of the phase boundary $H$ on the width of the deposit ring $W$.  The solid curve is the analytical result in the limit $\chi_i/p \to 0$; the other curves are the numerical results.  Different numerical curves correspond to different initial concentrations of the solute; values of parameter $\chi_i/p$ are shown at each curve.  The vertical scale is different from the horizontal scale by a factor of $\theta_i \ll 1$.}
\label{profileeps}
\end{center}
\end{figure}

However, the third and the fourth equations [Eqs.~(\ref{tilde-h-res})--(\ref{tilde-w-res})] represent entirely new results.  In the dimensional variables they yield the height of the phase boundary $H$ and the width of the deposit ring $W \equiv R_i - R$, respectively:
\begin{equation}
H(t) = \sqrt{\frac{\chi_i}p} \theta_i R_i \tilde H\left(\frac{t}{t_f}\right),
\label{height-result}\end{equation}
\begin{equation}
W(t) = \sqrt{\frac{\chi_i}p} R_i \tilde W\left(\frac{t}{t_f}\right),
\label{width-result}\end{equation}
where functions $\tilde H(\tau)$ and $\tilde W(\tau)$ are plotted in Figs.~\ref{resultseps}(c) and \ref{resultseps}(d) (the solid curves).  These results provide the sought dependence of the geometrical characteristics of the deposit ring on all the physical parameters of interest: on the initial geometry of the drop ($R_i$ and $\theta_i$), on the initial solute concentration ($\chi_i$), and on the time elapsed since the beginning of the drying process ($t$).  If the time is considered as a parameter, they can also be used to obtain the geometrical profile of the deposit ({\em i.e.}\ the dependence of the height on the width), which we plot by the solid line in Fig.~\ref{profileeps}.  Note that the vertical scale of this plot is highly expanded compared to the horizontal scale since there is an extra factor of $\theta_i \ll 1$ in the expression for the height; in the actual scale the height is much smaller than it appears in Fig.~\ref{profileeps}.

It is straightforward to obtain the asymptotics of $\tilde H(\tau)$ and $\tilde W(\tau)$ for early and late drying stages.  At early times, both the height and the width scale with the drying time as a power law with exponent $2/3$:
\begin{equation}
H \approx \sqrt{\frac{\chi_i}p} \theta_i R_i \frac{(3 \tau)^{2/3}}{2^{7/3}} \left[ 1 + O(\tau) \right]\qquad\qquad\left(\tau \ll 1\right),
\end{equation}
\begin{equation}
W \approx \sqrt{\frac{\chi_i}p} R_i \frac{(3 \tau)^{2/3}}{2^{7/3}} \left[ 1 + O(\tau) \right]\qquad\qquad\left(\tau \ll 1\right).
\end{equation}
Thus, at early times $H \approx \theta_i W$, which can also be deduced directly from Eq.~(\ref{constraint}) without obtaining the complete solution above.  [The early-time exponent $2/3$ was first obtained by Robert Deegan~\cite{deegan3} without deriving the full time dependence~(\ref{tilde-h-res})--(\ref{tilde-w-res}).]  At the end of the drying process, the height and the width approach finite values (which, apart from the dimensional scales, are universal, {\em i.e.}\ constants) and do so as power laws of $(t_f - t)$ with two different exponents:
\begin{equation}
H \approx \sqrt{\frac{\chi_i}p} \theta_i R_i \left[ \tilde H(1) - \frac{(1 - \tau)^{7/4}}{14 \tilde H(1)} + O(1 - \tau)^{5/2} \right]\qquad\quad\left(1 - \tau \ll 1\right),
\end{equation}
\begin{equation}
W \approx \sqrt{\frac{\chi_i}p} R_i \left[ \tilde W(1) - \frac{(1 - \tau)^{3/4}}{6 \tilde H(1)} + O(1 - \tau)^{3/2} \right]\qquad\quad\left(1 - \tau \ll 1\right),
\end{equation}
where $\tilde H(1)$ and $\tilde W(1)$ are simply numbers:
\begin{equation}
\tilde H(1) = \sqrt{\frac{1}3 {\rm B}\left(\frac{7}3,\frac{4}3\right)} \approx 0.297,
\end{equation}
\begin{equation}
\tilde W(1) = \int_0^1 \frac{1}{8 \tilde H(\tau)} \frac{\left[ 1 - (1 - \tau)^{3/4} \right]^{1/3}}{(1 - \tau)^{1/4}} \, d\tau \approx 0.609.
\end{equation}
Clearly, $dH/dW = \theta_i (1 - \tau)$ and hence vanishes when $\tau \to 1$.  This fact can also be observed in the flattening of the analytical graph in Fig.~\ref{profileeps} at late times.

Dependence of the height and the width on the radius of the drop $R_i$, while intuitively obvious (since $R_i$ is the only scale in this problem with the dimensionality of the length), has been verified in experiments~\cite{deegan3, deegan4}.  A linear fit has been obtained for the dependence of the ring width on the radius, in exact agreement with our findings.

\begin{figure}
\begin{center}
\includegraphics{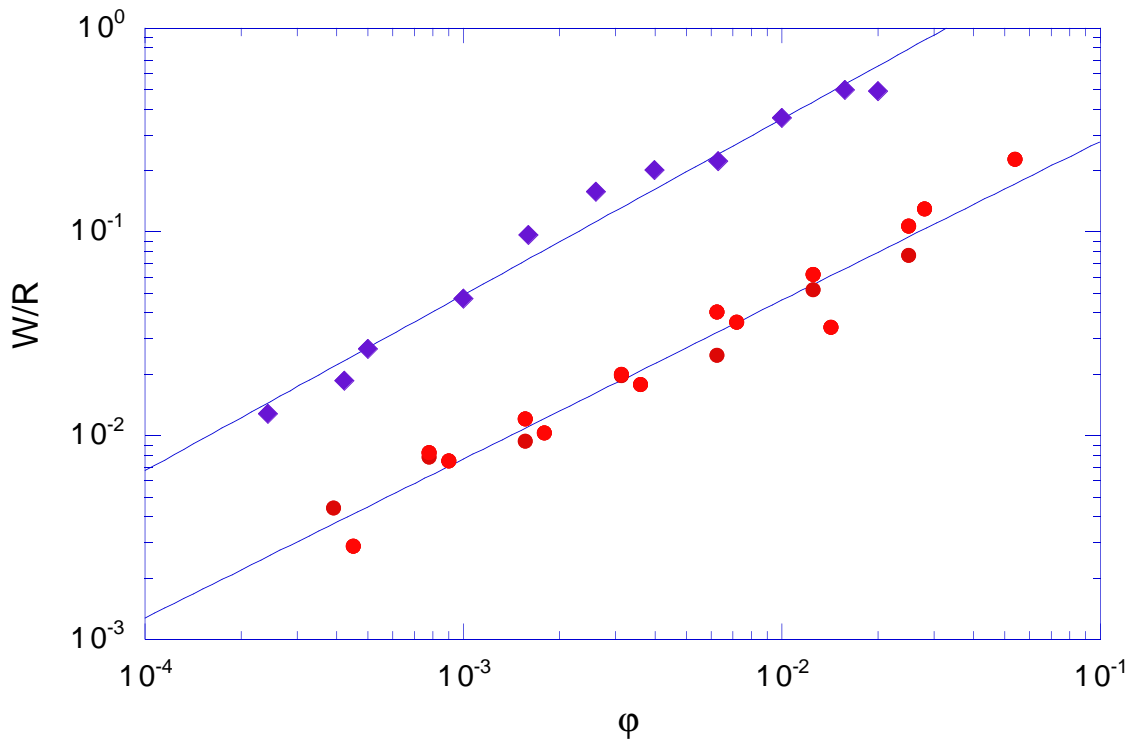}
\caption{Ring width normalized by the drop radius {\em vs.}\ initial concentration of the solute for two different particle sizes.  Experimental results, after Refs.~\cite{deegan3, deegan4}.  The two data sets are offset by a factor of 5 to avoid mixing of the data points related to the different particle sizes.  The lines running through the data are linear fits in the double-logarithmic scale, which upon conversion to the linear scale yield power laws with exponents $0.78 \pm 0.10$ and $0.86 \pm 0.10$.  (Courtesy Robert Deegan.)}
\label{widthconceps}
\end{center}
\end{figure}

\begin{figure}
\begin{center}
\includegraphics{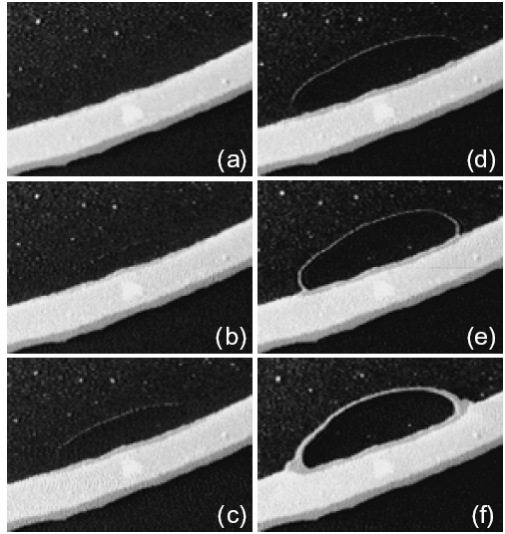}
\caption{A photographic sequence demonstrating a depinning event.  Experimental results, after Refs.~\cite{deegan3, deegan4}.  The view is from above, and the solid white band in the lower part of the frame is the ring; the rest of the drop is above the ring.  The time between the first and the last frames is approximately 6~s; the major axis of the hole is approximately 150~$\mu$m.  (Courtesy Robert Deegan.)}
\label{depinning}
\end{center}
\end{figure}

\begin{figure}
\begin{center}
\includegraphics{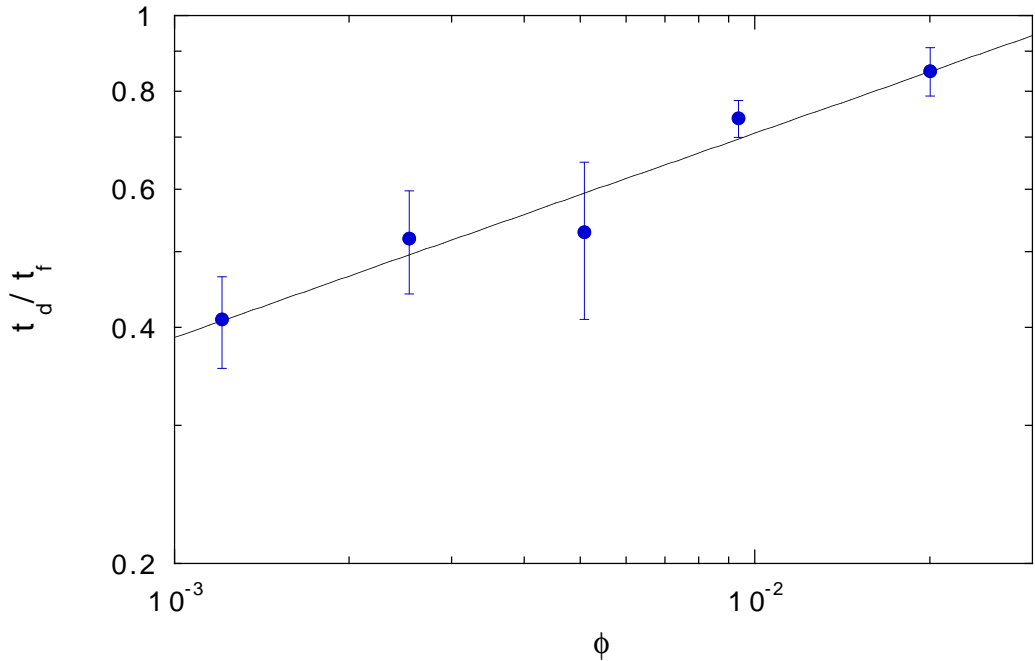}
\caption{Depinning time normalized by the extrapolated drying time {\em vs.}\ initial concentration of the solute.  Experimental results, after Refs.~\cite{deegan3, deegan4}.  The line running through the data is a linear fit in the double-logarithmic scale, which upon conversion to the linear scale yields a power law with exponent $0.26 \pm 0.08$.  (Courtesy Robert Deegan.)}
\label{timeconceps}
\end{center}
\end{figure}

Comparison to the experimental data for the dependence on the initial concentration of the solute is slightly less trivial.  Our results predict that both the height $H$ and the width $W$ scale with the initial concentration as $\chi_i^{1/2}$ (at least, in the leading order for small concentrations).  The same scaling prediction was also made by Robert Deegan~\cite{deegan3, deegan4}.  However, his experimental results show a different exponent of $\chi_i$: values $0.78 \pm 0.10$ and $0.86 \pm 0.10$ were obtained for two different particles sizes (Fig.~\ref{widthconceps}).  Why is the difference?  The answer lies in the fact that the width measured in the experiments~\cite{deegan3, deegan4} is not the full width of the ring at the end of the drying process, but rather the width of the ring at {\em depinning}.  Depinning is a process of the detachment of the liquid phase from the deposit ring (Fig.~\ref{depinning}).  This detachment was observed experimentally in colloidal suspensions but has not been explained in full theoretically yet.\footnote{While the full explanation is yet to be developed, the naive reason for the depinning seems relatively straightforward.  The {\em pinning\/} force depends only on the materials involved and is relatively insensitive to the value of the contact angle.  At the same time, the {\em depinning\/} force is simply the surface tension, which is directed along the liquid-air interface and which increases as the contact angle decreases (since only the horizontal component of this force is important).  Thus, the relatively constant pinning force cannot compensate for the increasing depinning force, and after the contact angle decreases past some threshold, the depinning force wins and causes the detachment.}  An important observation, however, is that the depinning time ({\em i.e.}\ the time at which the detachment occurs and the ring stops growing) depends on the initial concentration of the solute.  This dependence was also measured by Deegan (Fig.~\ref{timeconceps}), and the resulting exponent was determined to be $0.26 \pm 0.08$.  Thus, the width of the ring at depinning $W_d$ scales with the initial concentration of the solute $\chi_i$ as
\begin{equation}
W_d \propto \chi_i^{1/2} \tilde W\left(\frac{t_d}{t_f}\right) \propto \chi_i^{0.5} \tilde W\left(\chi_i^{0.26 \pm 0.08}\right),
\end{equation}
where $t_d$ is the depinning time ($t_d / t_f \propto \chi_i^{0.26 \pm 0.08}$).  As is apparent from Fig.~\ref{timeconceps}, the typical values of the depinning time are of the order of 0.4--0.8~$t_f$.  In this time range, function $\tilde W(t)$ is virtually linear [the analytical curve in Fig.~\ref{resultseps}(d)].  Therefore, the dependence of $W_d$ on $\chi_i$ has the overall exponent of the order of $0.76 \pm 0.08$.  It is now clear that both experimental results $0.78 \pm 0.10$ and $0.86 \pm 0.10$ fall within the range of the experimental uncertainty of this approximate predicted value, and the theoretical dependence of the ring width on the initial concentration agrees with the experimental results quite well.

Note that Robert Deegan~\cite{deegan3, deegan4} did {\em not\/} report direct measurements of the height of the deposit ring.  The height was {\em calculated\/} from the data in hand, and thus the direct comparison to the experimental data is not available for the height.

The square-root dependence of the height and the width on the concentration is in good agreement with general physical expectations.  Indeed, the volume of the deposit ring is roughly proportional to the product of the height and the width.  On the other hand, the height is of the same order of magnitude as the width since the ratio of the two is of the order of $\theta_i$ (which is a constant).  Thus, both the height and the width scale approximately as a square root of the ring volume.  Finally, the volume of the deposit ring is proportional to the initial volume fraction of the solute: the more solute is present initially, the larger the volume of the deposit ring is at the end.  Therefore, both the height and the width must scale as a square root of the initial volume fraction.  It is rewarding that our complex calculation leads to the same results as this simple physical argument.

Thus, the complete analytical solution to our model is available in the limit $\chi_i / p \to 0$, and this solution compares favorably with the experimental data.  Since the main-order solution in $\chi_i / p$ is perfectly adequate, the difference between the original system of equations and the one for the ``completely dry'' case is not important.  Indeed, the main-order results are identical in both cases, because one case in different from the other only by presence of $R$ instead of $R_i$ in a few places in the main equations, and this difference is of the correctional order in $\chi_i / p$.

\subparagraph{Numerical results for arbitrary initial concentrations of the solute.}  Apart from approaching the original system of equations~(\ref{constraint}), (\ref{main1}), (\ref{main2}), and (\ref{main3}) analytically, we also solve it numerically.  During this numerical procedure we do not presume that $\chi_i / p$ is small, nor do we expand any quantities or equations in $\epsilon$ or any other small parameters.  Our main purpose is to reproduce the results of the first part of this section and to determine the range of validity of our analytical asymptotics.

The typical values of $\chi_i / p$ in most experimental realizations are of the order of 0.001--0.01, and thus, only the concentrations below approximately 0.1 are of practical interest.  (Note that $\chi_i / p = 0.1$ corresponds to a quite substantial value of the small parameter $\epsilon = \sqrt{0.1} \approx 0.32$.)  Thus, we will concentrate on this range of $\chi_i / p$ when describing the results despite the fact the numerical procedure can be (and have been) conducted for any ratio $\chi_i / p$.  The general trend is illustrated well by the results in this range of concentrations.  In the case of $\chi_i$ comparable to $p$ our model is not expected to produce any sensible results, as the entire separation of the drop into the two phases (the liquid phase and the deposit phase) is based on the assumption that the mobility of the solute is qualitatively different in the two regions.  When $\chi_i$ is comparable to $p$ the two phases are physically indistinguishable, while the model still assumes they are different.

We present our numerical results for the same quantities (and in the same order) as in our analytical results~(\ref{theta-result}) and (\ref{fraction-result})--(\ref{width-result}).  Since for arbitrary $\chi_i / p$ time $t_f$ is {\em not\/} exactly the total drying time, there is a question of where (at what time) to terminate the numerical curves.  By convention, we terminate all the curves in all the graphs at value of $t / t_f$ when {\em all\/} the solute reaches the deposit phase.  In our model, it turns out that the time the last solute particles reach the deposit ring and the time the center-point of the liquid-air interface touches the substrate are about the same.  For all the initial concentrations, the two times were numerically found to be within 0.1\% of each other, and the curves are terminated at exactly this moment.  Of course, in reality a small fraction of solute should stay in the liquid phase as long as the liquid phase exists, and so the moment the last solute particles reach the deposit phase should be {\em after\/} the moment the center-point touches the substrate; however, the amount of solute remaining in the liquid phase at touchdown is insignificant, and practically all the deposit has already formed.

Numerical results for angle $\theta$ as a function of time are shown in Fig.~\ref{resultseps}(a).  All curves behave almost linearly (as expected), however, the slope increases with concentration: formation of the ring in the drops with more solute finishes faster (in the relative scale of $t_f$).  The end of each curve demonstrates the value of the angle $\theta_t$ at the moment the liquid-air interface touches the substrate.  The analytical expression for this angle is $\theta_t = - 2 H / R$ for a thin drop.  The absolute value of this angle increases with concentration, which is quite natural since for small concentrations the height of the ring grows as a square root of the concentration while the radius of the liquid phase does not change substantially.  Clearly, the numerical results converge to the analytical curve when $\chi_i / p \to 0$.

Growth of the volume fraction of solute in the deposit phase $V_D^S/V^S$ with time is shown in Fig.~\ref{resultseps}(b) for various solute concentrations.  This graph reconfirms the observation of the preceding paragraph that the solute transfer happens faster (in units of $t_f$) for denser colloidal suspensions.  All curves are terminated when volume fraction $V_D^S/V^S$ becomes equal to 1.  (The apparent termination of the curve for $\chi_i / p = 0.1$ earlier than that is an artifact of the plotting software.)  As the corresponding analytical results do, the numerical plots of Figs.~\ref{resultseps}(a) and \ref{resultseps}(b) should presumably hold true independently of the geometrical details of the solute accumulation in the ring (which cannot be expected from the following plots for the ring height and width).

The next two graphs represent the numerical results for the height [Fig.~\ref{resultseps}(c)] and the width [Fig.~\ref{resultseps}(d)] of the deposit ring as functions of time.  The ring profile, {\em i.e.}\ the dependence of the height on the width, is also shown in Fig.~\ref{profileeps}.  As the graphs depict, the ring becomes wider and lower (in the reduced variables) for higher initial concentrations of the solute.  Since the volume of the ring is roughly proportional to the product of the height and the width, the decrease in height must be of the same magnitude as the increase in width.  This can be qualitatively observed in the graphs.

\begin{figure}
\begin{center}
\includegraphics{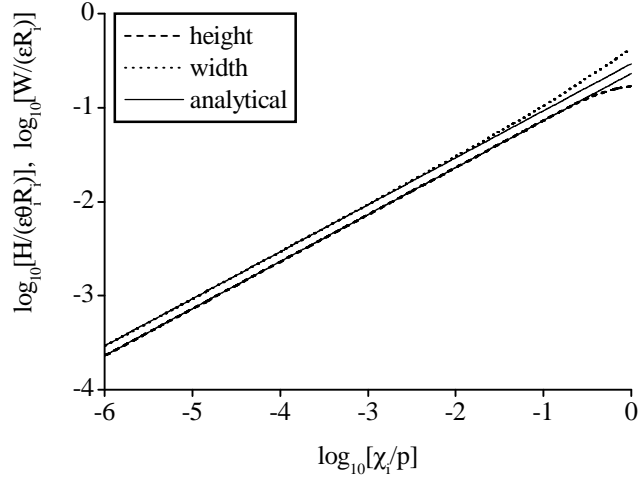}
\caption{Numerical results: log-log plot of the dependence of the height of the phase boundary $H$ and the width of the deposit ring $W$ on the initial volume fraction of the solute $\chi_i$.  The main-order analytical results $H \propto \sqrt{\chi_i/p}$ and $W \propto \sqrt{\chi_i/p}$ are also provided for comparison.}
\label{conceps}
\end{center}
\end{figure}

As a final piece of the numerical results, we create a double-logarithmic plot for the dependence of the height and the width on the initial concentration of the solute (Fig.~\ref{conceps}).  The predicted square-root dependence on the initial concentration is seen to hold true for volume fractions up to approximately $10^{-1/2} p$ for the height and up to approximately $10^{-3/2} p$ for the width.  The deviations for higher volume fractions are due to the increasing role of the correctional terms in $\epsilon$ compared to the main-order terms represented by the solid lines.  In this graph, as in all the results of this section, it is clear that our main-order analytical results provide an adequate description of all the functional dependencies in the range of the initial concentrations of experimental importance (0.001--0.01).

In general, our numerical results complement and reinforce our analytical results, providing a crosscheck of both methods.

\section{Discussion}

Both the analytical results of Eqs.~(\ref{theta-result}) and (\ref{fraction-result})--(\ref{width-result}) and the numerical graphs of Figs.~\ref{resultseps} and \ref{profileeps} may be reproduced experimentally giving validation to the proposed model.  Measurements of the profiles in Fig.~\ref{profileeps} should be particularly easy to conduct (since there is no time dependence involved) and may confirm or refute the predicted robustness and the universality of the deposition profiles.

While the main principles of the proposed model were laid down by Robert Deegan~\cite{deegan3, deegan4}, its analytical solution for small concentrations and its numerical solution for arbitrary concentrations are obtained here for the first time.  The availability of the exact analytical solution demonstrated that the theoretical scaling of the deposit width at depinning with solute concentration indeed agrees with both measured values of the exponents.  The earlier estimate of Refs.~\cite{deegan3, deegan4} based solely on the early-time exponent had an overlap with only one of the concentration exponents.  Our numerical results also quantify the range of solute concentrations where the predicted square-root dependence of the width holds true.  In general, Robert Deegan's results were not sufficient to obtain the proper scaling of the width with time anywhere beyond the very early drying stages; the results of this work provide that time scaling at {\em all\/} drying stages.  All the presented results suggest that the deposit ring profile and its growth can be fully accounted for on the basis of the finite volume of the solute particles only and that the governing functional dependences are {\em universal.}

One may notice that the curves in Fig.~\ref{profileeps} end at some positive (non-zero) height.  This indicates the solute is exhausted before the profile curves have a chance to return to the substrate, and the final shape of the deposit ring must have a vertical wall at its inner side.  We believe this is an artifact of our model, which is inherently two-dimensional when flows inside the drop are concerned.  Thus, the vertical distribution of the solute was assumed homogeneous (the phase boundary is vertical and the particles get stacked uniformly at all heights), and we used the depth-averaged velocity~(\ref{defv}) throughout this work.  This is equivalent to assuming that vertical mixing is complete.  This assumption is quite important, and the results are expected to get modified if the true three-dimensional velocity profile is used instead of the depth-averaged velocity.  We expect that if a three-dimensional model were built and the dependence on $z$ were taken into account for all the quantities then the discontinuous wall of the phase boundary would get smoothened and the height would continuously return to zero.  A question remains whether such a model would be solvable analytically.

Our model relies on the assumption that solute mobility is different in the so-called liquid and deposit phases.  In essence, we assume that the mobility is 0 in the deposit phase and 1 in the liquid phase.  This assumption, while artificial in its nature, seems relatively reasonable when applied to this system.  Indeed, in the physical situations near the close packing, the loss of mobility typically occurs over a quite narrow range of the concentration values, and hence our assumption should work satisfactorily when the difference between $\chi_i$ and $p$ is in the orders of magnitude.  The higher the initial concentration is and the closer the two values are, the worse this assumption holds true and the more artificial the difference between the two phases is.  Thus, the validity of any model based on this separation of the mobility scales decreases for higher initial concentrations of the solute.

The model assumes that the free-surface slope between the liquid and the deposit phases is continuous.  In fact, assumption~(\ref{constraint}) expressing this continuity is one of the four basic equations of this work.  This assumption seems quite natural as well.  Indeed, if the liquid is present on both sides of the phase boundary, the change in the slope of its free surface would cost extra energy from the extra curvature at the phase boundary, since the liquid-air interface possesses effective elasticity.  Presence of this extra energy (or the extra pressure) at the location of the phase boundary is not justified by any physical reasons as all the processes are slow and the surface is in equilibrium.  In equilibrium, the surface shape must have constant curvature past the phase boundary since the entire separation into the two phases is quite artificial as discussed above.  Presence of the particles below the liquid-air interface does not influence the surface tension, and thus the liquid surface (and its slope) should be continuous at the phase boundary.  If the density of the particles matches the density of the liquid (which was the case in the experiments), nothing prevents the particles from filling up the entire space between the substrate and the liquid-air interface, thus providing the growth of the upper edge of the deposit phase {\em along\/} the liquid-air interface.  This is particularly true for the thin drops discussed here, where vertical mixing is intensive, where the free surface is nearly horizontal, and where the problem is essentially two-dimensional.  However, the equality of the slopes on both sides of the phase boundary does {\em not\/} seem inevitable, and one may think of the situations when it does get violated.  One example might be the late drying times, when the deposit growth is very fast [Fig.~\ref{resultseps}(b)] and hence the deposition may occur in some non-regular manner inconsistent with this slow-process description.  Other examples may be related to gravity (slightly unequal densities of the particles and the fluid) or convection.  This assumption can possibly be checked experimentally, and if condition~(\ref{constraint}) is found violated, an equivalent constraint dependent on the details of the deposit-growth mechanism must be constructed in place of Eq.~(\ref{constraint}).

Another inherent assumption of our model is related to the evaporation rate $J(r)$.  Presence of the solute inside the drop was assumed not to affect the evaporation from its surface.  This is generally true when the evaporation is not too fast and the deposit phase is not too thick and not too concentrated.  When these conditions are not obeyed, presence of a thick or concentrated layer of the solute on the way of the liquid moving from the phase boundary to the contact line may create a strong viscous force.  This viscous force would prevent the necessary amount of fluid from being supplied to the intensive-evaporation region near the contact line.  Generally, we assumed throughout this work that the viscous stresses are not important, and this is valid whenever $v \ll \sigma / 3 \eta$.  In the deposit phase, the velocity is large due to the proximity to the contact-line divergence of the evaporation rate, and the effective viscosity is large due to the high concentration of the solute.  Thus, this condition may get violated and the viscosity may become important in the deposit phase, slowing down the supply of the liquid and ultimately making the deposit dry.  Obviously, this affects the evaporation rate, and the functional form of the evaporation profile changes.  Simple assumption that the evaporation rate stays of the same functional form, but with the divergence at the phase boundary (at $R$) instead of the contact line (at $R_i$), was shown above not to affect our main-order results.  Thus, our results appear to be relatively insensitive to the exact location of this divergence within the (narrow) deposit phase.  (In reality the evaporation edge would be somewhere between the original contact line and the phase boundary, {\em i.e.}\ the real situation is intermediate between the two considered.)  However, the deposit could modify the evaporation rate $J(r)$ in other ways.  When there is a dry deposit ring just outside the liquid phase, the entire functional form of $J$ may change, and the Laplace equation for an equivalent electrostatic problem must be solved anew with additional boundary conditions responsible for the presence of the dry solute rim and the modified evaporation at the edge.  As we already saw in the Appendix, this is the most complicated part of the problem, and the mathematics can become prohibitively complex.  Thus, finding the exact form of $J$ may be a formidable task.  One way around is in creating such evaporating conditions that the functional profile is simpler, for instance, $J$ is just a constant.  This would be more difficult to control experimentally, but would be much easier to treat analytically.  The unavailability of the exact analytical form for $J$ seems to be the biggest open question in this class of problems~\cite{popov2, popov4}.

The equilibrium surface shape of the liquid phase is a spherical cap~(\ref{h-finite}).  This is a rigorous result valid during most of the drying process.  However, when $h(0,t)$ becomes negative and exceeds $H(t)$ in its absolute value ({\em i.e.}\ when the center-point touches the substrate), the surface shape is no longer spherical.  Moreover, a new element of the contact line is introduced in the center of the drop in addition to the original contact line at the perimeter, and the entire evaporation profile gets modified in addition to the modified surface shape, thus influencing all the other quantities.  Our treatment does not account for the small fraction of the drying process occurring after this touchdown (which is a change in topology of the free surface, and thus requires a separate treatment after it happened).  First of all, the amount of liquid remaining in the drop at this moment is of the order of $\epsilon$ compared to the original volume, and hence it would not modify our main-order analytical results.  Second, as our {\em numerical\/} calculations show, at touchdown practically all the solute is already in the deposit phase, and the remaining amount of solute in the liquid phase is insignificant.  Thus, within our model, the remainder of the drying process cannot modify the deposit ring substantially, and hence this neglect of the late-time regime seems well justified.  Experimentally, the inner part of the deposit ring is different from our prediction (which is a vertical wall) and appears to have a spread shelf.  Presence of this tail in the deposit distribution can be caused by several features absent in our model.  Its inherent two-dimensionality may be one of these shortcomings (as discussed above); the account for the dynamical processes occurring after the deposit phase has already been formed ({\em e.g.}\ avalanches of the inner wall) may be another missing feature.  Absence of the treatment of the late-time regime may be among these reasons influencing the final distribution of the deposit as well.  A more detailed account for the effects of this late-time regime might be required in the future.

\vspace{3ex}

{\small  This work was completed as a part of the Ph.D.\ dissertation research supervised by Thomas A.\ Witten.  The author acknowledges valuable input from Todd F.\ Dupont and Robert R.\ Deegan.  This work was supported in part by the National Science Foundation MRSEC Program under award number DMR-0213745.}

\section*{Appendix: On the evaporation rate}

The purpose of this section is to obtain the evaporation rate from the free surface of a round sessile drop on the substrate.  Since presence of the solute is irrelevant to this purpose (at least for its low concentrations), one may assume that the solute is simply absent and the drop is just pure water.  We first consider the generic problem with an arbitrary contact angle $\theta$, and then find the appropriate limit of interest $\theta \ll 1$.  If the radius of the drop footprint on the substrate is $R_i$, then its surface shape for an arbitrary contact angle is given by
\begin{equation}
h(r,t) = \sqrt{\frac{R_i^2}{\sin^2 \theta(t)} - r^2} - R_i \cot\theta(t).
\end{equation}
Despite the fact this problem is two centuries old, some results are presented here in their closed analytical form for the first time, and some correct earlier expressions.

Our task involves solution of the equivalent electrostatic problem (the Laplace equation) for the conductor of the shape of the drop plus its reflection in the plane of the substrate (kept at constant potential, as a boundary condition).  In the case of the round drop the shape of this conductor resembles a symmetrical double-convex lens comprised of two spherical caps.  The system of orthogonal coordinates that matches the symmetry of this object (so that one of the coordinate surfaces coincides with the surface of the lens) is called the toroidal coordinates $(\alpha,\beta,\phi)$, where coordinates $\alpha$ and $\beta$ are related to the cylindrical coordinates $r$ and $z$ by
\begin{equation}
r = \frac{R_i \sinh\alpha}{\cosh\alpha - \cos\beta},\qquad\qquad z = \frac{R_i \sin\beta}{\cosh\alpha - \cos\beta},
\end{equation}
and the azimuthal angle $\phi$ has the same meaning as in the cylindrical coordinates.  Solution to the Laplace equation in the toroidal coordinates involves the Legendre functions of fractional degree and was derived in a book by Lebedev~\cite{lebedev}.  The electrostatic potential or vapor density is independent of the azimuthal angle $\phi$ and reads
$$ n(\alpha,\beta) = n_\infty + (n_s - n_\infty) \sqrt{2(\cosh\alpha - \cos\beta)} \times$$
\begin{equation}
\times \int_0^\infty \frac{\cosh\theta\tau \cosh(2\pi - \beta)\tau}{\cosh\pi\tau \cosh(\pi - \theta)\tau} P_{-1/2 + i\tau}(\cosh\alpha) \, d\tau.
\end{equation}
Here $n_s$ is the density of the saturated vapor just above the liquid-air interface (or the potential of the conductor), $n_\infty$ is the ambient vapor density (or the value of the potential at infinity), and $P_{-1/2 + i\tau}(x)$ are the Legendre functions of the first kind (they are real valued).  The surface of the lens is described by the two coordinate surfaces $\beta_1 = \pi - \theta$ and $\beta_2 = \pi + \theta$, and the $\beta$ derivative is normal to the surface.  The evaporation rate from the surface of the drop is therefore given by
\begin{equation}
J(\alpha) = D \frac{1}{h_\beta} \left.\partial_\beta n(\alpha,\beta)\right|_{\beta = 2\pi + \beta_1} = D \frac{\cosh\alpha - \cos\beta}{R_i} \left.\partial_\beta n(\alpha,\beta)\right|_{\beta = 3\pi - \theta},
\end{equation}
where $D$ is the diffusion constant and $h_\beta = R_i /(\cosh\alpha - \cos\beta)$ is the metric coefficient in coordinate $\beta$.  (Note that an incorrect expression for $J$ with a plus sign in the metric coefficient was used in Eq.~(A2) of Ref.~\cite{deegan1}.)  Thus, the exact analytical expression for the absolute value of the evaporation rate as a function of $r$ is available:
$$J(r) = \frac{D(n_s - n_\infty)}{R_i} \left[\frac{1}2 \sin\theta + \sqrt{2} \left(\cosh\alpha + \cos\theta\right)^{3/2} \times \right.$$
\begin{equation}
\left. \times \int_0^\infty \frac{\cosh\theta\tau}{\cosh\pi\tau} \tanh\left[(\pi - \theta)\tau\right] P_{-1/2 + i\tau}(\cosh\alpha) \, \tau d\tau\right],
\label{j-circular}\end{equation}
where the toroidal coordinate $\alpha$ is uniquely related to the polar coordinate $r$ on the surface of the drop:
\begin{equation}
r = \frac{R_i \sinh\alpha}{\cosh\alpha + \cos\theta}.
\end{equation}
Expression~(\ref{j-circular}) is valid for an {\em arbitrary\/} contact angle $\theta$ and corrects an earlier expression of Ref.~\cite{hu} [Eq.~(28)] where a factor of $\sqrt{2}$ in the second term inside the square bracket is missing.

The expression for the evaporation rate is not operable analytically in most cases, as it represents an integral of a non-trivial special function (which, in its turn, is an integral of some simpler elementary functions).  In most cases, it is necessary to recourse to the asymptotic expansions in the contact angle $\theta$ in order to obtain any meaningful analytical expressions.  However, there is one exception to this general statement.  An important quantity is the total rate of water mass loss by evaporation $dM/dt$, which sets the time scale for all the processes.  This total rate can be expressed as an integral of the evaporation rate (defined as the evaporative mass loss per unit surface area per unit time) over the surface of the drop:
\begin{equation}
\frac{dM}{dt} = - \int_A J(r) \sqrt{1+(\nabla h)^2} \, r dr d\phi = - \int_0^{R_i} J(r) \sqrt{1 + (\partial_r h)^2} \, 2 \pi r dr,
\label{dmdt-circular}\end{equation}
where the first integration is over the substrate area $A$ occupied by the drop.  This expression actually involves triple integration: one in the expression above as an integral of $J(r)$, another in the expression for $J(r)$ as an integral of the Legendre function of the first kind, and the third as an integral representation of the Legendre function in terms of the elementary functions.  However, it is possible to simplify the above expression significantly and reduce the number of integrations from three to one.  Investing some technical effort and using Eq.~(2.17.1.10) of Ref.~\cite{prudnikov}, one can obtain a substantially simpler result that does not involve any special functions at all:
$$\frac{dM}{dt} = - \pi R_i D (n_s - n_\infty) \left[\frac{\sin\theta}{1+\cos\theta} + \right.$$
\begin{equation}
\left. + 4 \int_0^\infty \frac{1 + \cosh 2\theta\tau}{\sinh 2\pi\tau} \tanh\left[(\pi - \theta)\tau\right] \, d\tau \right]
\label{dmdtexact}\end{equation}
(not reported in the literature previously).  This result together with the expression for the total mass of water
\begin{equation}
M = \rho \int_0^{R_i} h(r,t) \, 2 \pi r dr = \pi \rho R_i^3 \frac{\cos^3 \theta - 3 \cos\theta + 2}{3 \sin^3 \theta}
\label{watermass-circular}\end{equation}
(where $\rho$ is the water density) provides a direct method for finding the time dependence of $\theta$ for an {\em arbitrary\/} value of the contact angle.  Combining the time derivative of the last expression with result~(\ref{dmdtexact}), one can obtain a single differential equation for $\theta$ as a function of time $t$:
$$\frac{d\theta}{dt} = - \frac{D (n_s - n_\infty)}{\rho R_i^2} (1+\cos\theta)^2 \left[\frac{\sin\theta}{1+\cos\theta} + \right.$$
\begin{equation}
\left. + 4 \int_0^\infty \frac{1 + \cosh 2\theta\tau}{\sinh 2\pi\tau} \tanh\left[(\pi - \theta)\tau\right] \, d\tau \right].
\label{dthetadtexact}\end{equation}
Having determined the dependence $\theta(t)$ from this equation, one can obtain the time dependence of any other quantity dependent on the contact angle, for instance, the time dependence of the mass from relation~(\ref{watermass-circular}), or any other geometrical quantity considered earlier.

In practice, however, the analytical calculations in a closed form cannot be conducted any further for arbitrary contact angles, and we will use the limit of small contact angles in all the subsequent analytical calculations.  Besides being the limit of our interest and most practical importance, this limit is also perfectly adequate even for quite substantial angles, as will be seen in a moment.

Expanding the right-hand side of Eq.~(\ref{dthetadtexact}) in small $\theta$, we immediately obtain that the contact angle decreases {\em linearly\/} with time in the main order of this expansion:
\begin{equation}
\theta = \theta_i \left( 1 - \frac{t}{t_f} \right),
\label{theta-circular}\end{equation}
where we introduced the total drying time $t_f$ defined in terms of the initial contact angle $\theta_i = \theta(0)$:
\begin{equation}
t_f = \frac{\pi \rho R_i^2 \theta_i}{16 D (n_s - n_\infty)}.
\label{dryingtime}\end{equation}
In the main order, the total rate of water mass loss is constant and the water mass also decreases with time linearly:
\begin{equation}
M = \frac{\pi \rho R_i^3 \theta_i}4 \left( 1 - \frac{t}{t_f} \right).
\label{massofwater}\end{equation}
This linear time dependence during the vast majority of the drying process was directly confirmed in the experiments~\cite{deegan3, deegan4}; see Fig.~\ref{thetatimeeps}.  The dependence of the evaporation rate~(\ref{dmdtexact}) on radius (linearity in $R_i$) was also confirmed experimentally and is known to hold true for the case of the diffusion-limited evaporation~\cite{davies}.

\begin{figure}
\begin{center}
\includegraphics{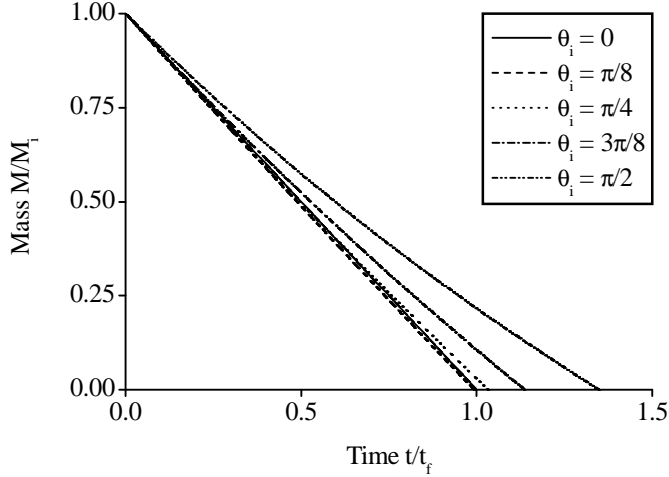}
\caption{Numerical results: dependence of water mass $M$ on time $t$.  Different curves correspond to different initial contact angles; values of parameter $\theta_i$ are shown at each curve.  The analytical result [Eq.~(\ref{massofwater})] in limit $\theta_i \to 0$ is also provided (the solid curve).}
\label{m-teps}
\end{center}
\end{figure}

In Fig.~\ref{m-teps}, we plot the exact numerical solution for $M(t)$ based on Eqs.~(\ref{dthetadtexact}) and (\ref{watermass-circular}) for several values of the initial contact angle $\theta_i$ together with the small-angle asymptotic of Eq.~(\ref{massofwater}).  In this figure, $M_i$ is the initial mass of water in the drop defined by the pre-factor in Eq.~(\ref{massofwater}).  [Note that $t_f$ is {\em not\/} the total drying time for each $\theta_i$; instead, it is just the combination of the problem parameters defined in Eq.~(\ref{dryingtime}), which coincides with the total drying time only when $\theta_i \to 0$.]  Fig.~\ref{m-teps} demonstrates that the small-angle approximation works amazingly well up to the angles as large as 45~degrees, and therefore, no precision or generality is lost by working in the limit of small contact angles for the typical experimental values of $\theta_i$.  Lastly, we note that the large-angle corrections may be responsible for the observed non-linearity of the experimentally measured dependence $M(t)$, as is clear from the comparison of Fig.~\ref{m-teps} (theory) and Fig.~\ref{thetatimeeps} (experiment).

Expression for the evaporation rate~(\ref{j-circular}) becomes particularly simple in the limit of small contact angles.  Employing one of the integral representations of the Legendre function in terms of the elementary functions (Eq.~(7.4.7) of Ref.~\cite{lebedev}), it is relatively straightforward to obtain the following result:
\begin{equation}
J(r) = \frac{D (n_s - n_\infty)}{R_i} \frac{2}\pi \cosh\frac{\alpha}2\qquad\qquad(\theta \to 0),
\end{equation} 
which, upon identification $\cosh\alpha = (R_i^2 + r^2)/(R_i^2 - r^2)$ for $\theta = 0$, can be further reduced to Eq.~(\ref{evaprate-circular}).  Thus, for thin drops the expression for the evaporation rate reduces to an extremely simple result featuring the one-over-the-square-root divergence near the edge of the drop.  The same result could have been obtained directly if we solved an equivalent electrostatic problem for an infinitely thin disk instead of the double-convex lens.  It is particularly rewarding that after all the laborious calculations the asymptotic of our result is in exact agreement with the predictions of a textbook (see Ref.~\cite{jackson} for the derivation of the one-over-the-square-root divergence of the electric field near the edge of a conducting plane in the three-dimensional space).  Eq.~(\ref{evaprate-circular}) is the result we were looking for in our case of the thin circular drops.

For the sake of completeness, it is also interesting to note the opposite limit of the expression~(\ref{j-circular}), when the surface of the drop is a hemisphere ($\theta = \pi / 2$).  In this limit, a similar calculation can be conducted, and the uniform evaporation rate is recovered:
\begin{equation}
J(r) = \frac{D (n_s - n_\infty)}{R_i}\qquad\qquad(\theta \to \pi/2).
\end{equation}
This result is also in perfect agreement with the expectations; the same result could have been obtained if we directly solved the Laplace equation for a sphere (the hemispherical drop and its reflection in the substrate).  The uniform evaporation rate is a result of the full spherical symmetry of such a system.  Similar exact results can also be obtained for a few other discrete values of the contact angle ({\em e.g.}\ for $\theta = \pi/4$).



\begin{thebibliography}{99}

\bibitem{pre1} N.D.~Denkov, O.D.~Velev, P.A.~Kralchevsky, I.B.~Ivanov, H.~Yoshimura, K.~Nagayama, {\em Langmuir\/} {\bf 8}, 3183 (1992).

\bibitem{pre2} A.S.~Dimitrov, C.D.~Dushkin, H.~Yoshimura, K~Nagayama, {\em Langmuir\/} {\bf 10}, 432 (1994).

\bibitem{dushkin} C.D.~Dushkin, H.~Yoshimura, K.~Nagayama, {\em Chem.\ Phys.\ Lett.} {\bf 204}, 455 (1993).

\bibitem{pre3} T.~Ondarcuhu, C.~Joachim, {\em Europhys.\ Lett.} {\bf 42}, 215 (1998).

\bibitem{pre4} J.~Boneberg, F.~Burmeister, C.~Shafle, P.~Leiderer, D.~Reim, A.~Fery, S.~Herminghaus, {\em Langmuir\/} {\bf 13}, 7080 (1997).

\bibitem{jpcb2} R.G.~Larson, T.T.~Perkins, D.E.~Smith, S.~Chu, {\em Phys.\ Rev.\ E\/} {\bf 55}, 1794 (1997).

\bibitem{jpcb1} J.P.~Jing, J.~Reed, J.~Huang, X.~Hu, V.~Clarke, J.~Edington, D.~Housman, T.S.~Anantharaman, E.J.~Huff, B.~Mishra, B.~Porter, A.~Shenkeer, E.~Wolfson, C.~Hiort, R.~Kantor, C.~Aston, D.C.~Schwartz, {\em Proc.\ Natl.\ Acad.\ Sci.\ U.S.A.} {\bf 95}, 8046 (1998).

\bibitem{pre5} A.B.~El~Bediwi, W.J.~Kulnis, Y.~Luo, D.~Woodland, W.N.~Unertl, {\em Mater.\ Res.\ Soc.\ Symp.\ Proc.} {\bf 372}, 277 (1995).

\bibitem{pre6} F.~Parisse, C.~Allain, {\em J.\ Phys. II\/} {\bf 6}, 1111 (1996).

\bibitem{pre7} F.~Parisse, C.~Allain, {\em Langmuir\/} {\bf 13}, 3598 (1996).

\bibitem{pre8} E.~Adachi, A.S.~Dimitrov, K.~Nagayama, in {\em Film Formation in Waterborne Coatings,} edited by T.~Provder, M.A.~Winnik, M.W.~Urban, p. 419 (American Chemical Society, Washington DC, 1996).

\bibitem{pre9} E.~Adachi, A.S.~Dimitrov, K.~Nagayama, {\em Langmuir\/} {\bf 11}, 1057 (1995).

\bibitem{shmuylovich} L.~Shmuylovich, A.Q.~Shen, H.A.~Stone, {\em Langmuir\/} {\bf 18}, 3441 (2002).

\bibitem{pre0} J.~Conway, H.~Korns, M.R.~Fisch, {\em Langmuir\/} {\bf 13}, 426 (1997).

\bibitem{jpcb3} K.S.~Birdi, D.T.~Vu, A.~Winter, {\em J.\ Phys.\ Chem.} {\bf 93}, 3702 (1989).

\bibitem{jpcb4} K.S.~Birdi, D.T.~Vu, {\em J.\ Adhes.\ Sci.\ Technol.} {\bf 7}, 485 (1993).

\bibitem{jpcb5} M.E.R.~Shanahan, C.~Bourges, {\em Int.\ J.\ Adhes.} {\bf 14}, 201 (1994).

\bibitem{jpcb6} C.~Bourges, M.E.R.~Shanahan, {\em Langmuir\/} {\bf 11}, 2820 (1995).

\bibitem{jpcb7} S.M.~Rowan, G.~McHale, M.I.~Newton, {\em J.\ Phys.\ Chem.\ B\/} {\bf 99}, 13268 (1995).

\bibitem{jpcb8} S.M.~Rowan, G.~McHale, M.I.~Newton, M.~Toorneman, {\em J.\ Phys.\ Chem.\ B\/} {\bf 101}, 1265 (1997).

\bibitem{deegan1} R.D.~Deegan, O.~Bakajin, T.F.~Dupont, G.~Huber, S.R.~Nagel, T.A.~Witten, {\em Phys.\ Rev.\ E\/} {\bf 62}, 756 (2000).

\bibitem{deegan2} R.D.~Deegan, O.~Bakajin, T.F.~Dupont, G.~Huber, S.R.~Nagel, T.A.~Witten, {\em Nature\/} {\bf 389}, 827 (1997).

\bibitem{deegan3} R.D.~Deegan, {\em Phys.\ Rev.\ E\/} {\bf 61}, 475 (2000).

\bibitem{deegan4} R.D.~Deegan, {\em Ph.D.\ thesis\/} (University of Chicago, Dept.\ of Physics, 1998).

\bibitem{lebedev} N.N.~Lebedev, {\em Special Functions and Their Applications,} Revised English ed., Chapters 7 and 8 (Prentice-Hall, Englewood Cliffs, 1965).

\bibitem{jpcb0} R.G.~Picknett, R.~Bexon, {\em J.\ Colloid Interface Sci.} {\bf 61}, 336 (1977).

\bibitem{hu} H.~Hu, R.G.~Larson, {\em J.\ Phys.\ Chem.\ B\/} {\bf 106}, 1334 (2002).

\bibitem{popov1} Y.O.~Popov, T.A.~Witten, {\em Eur.\ Phys.\ J.\ E\/} {\bf 6}, 211 (2001).

\bibitem{popov2} Y.O.~Popov, T.A.~Witten, {\em Phys.\ Rev.\ E\/} {\bf 68}, 036306 (2003).

\bibitem{dupont} T.~Dupont, private communication.

\bibitem{popov4} Y.O.~Popov, {\em Ph.D.\ thesis\/} (University of Chicago, Dept.\ of Physics, 2003), {\tt cond-mat/0312196}.

\bibitem{popov3} Y.O.~Popov, {\em J.\ Colloid Interface Sci.} {\bf 252}, 320 (2002).

\bibitem{prudnikov} A.P.~Prudnikov, Y.A.~Brychkov, O.I.~Marichev, {\em Integrals and Series,} Vol. 3 (Gordon and Breach, London, 1986).

\bibitem{davies} J.T.~Davies, E.K.~Rideal, {\em Interfacial Phenomenon\/} (Academic Press, New York, 1963).

\bibitem{jackson} J.D.~Jackson, {\em Classical Electrodynamics,} 2nd ed., Chapters 2 and 3 (Wiley, New York, 1975).

\end{thebibliography}
\end{document}